\pdfoutput=1
\documentclass[aps,prb,superscriptaddress,reprint,floatfix,nobalancelastpage,nofootinbib]{revtex4-2}
\usepackage{import}
\usepackage{url}
\usepackage{csquotes}
\usepackage{ifthen}

\usepackage{refcount, fmtcount}

\usepackage[T1]{fontenc}

\usepackage{shellesc}

\usepackage{import}

\usepackage{stmaryrd} \usepackage{amsmath}
\usepackage{amssymb}
\usepackage{amstext}
\usepackage{scalerel} \usepackage{mathtools}
\usepackage{subdepth} 

\usepackage{siunitx} \renewenvironment{subequations}{\begin{defaultsubequations}
		
		\renewcommand\notag{}
		}{\end{defaultsubequations}
}

\usepackage{graphicx}
\usepackage[space]{grffile} \usepackage[caption=false,farskip=0pt]{subfig}
\usepackage[export]{adjustbox}

\usepackage{color}
\usepackage{colortbl}
\usepackage{xcolor} \definecolor{myblue}{HTML}{0072BD}
\definecolor{myred}{HTML}{A2142F}
\definecolor{myorange}{HTML}{D95319}
\definecolor{myviolet}{HTML}{7E2F8E}
\definecolor{mygreen}{HTML}{77AC30}
\definecolor{mylightblue}{HTML}{4DBEEE}
\definecolor{myyellow}{HTML}{EDB120}

\definecolor{mylightgray}{HTML}{E0E0E0}
\definecolor{myverylightgray}{HTML}{EEEEEE}
\definecolor{mymiddlegray}{HTML}{D0D0D0}
\definecolor{mydarkgray}{HTML}{CCCCCC}
\definecolor{mydarkergray}{HTML}{888888}
\definecolor{myblack}{HTML}{000000}
\definecolor{mywhite}{HTML}{FFFFFF}

\usepackage{multirow} \usepackage{array}

\usepackage{booktabs} 
\let\defaultbottomrule\bottomrule
\renewcommand{\toprule}{\defaultbottomrule\defaultbottomrule\addlinespace[2pt]}
\renewcommand{\bottomrule}{\defaultbottomrule\defaultbottomrule}

\usepackage[
	linktocpage=true,
	plainpages=false,
	hypertexnames=false,
	pdfauthor={Stanislav Kazmin},
	pdftitle={},
	pdfkeywords={},
	pdfcreator={LuaLaTeX},
	pdfproducer={LuaLaTeX},
	pdfsubject={},
	unicode=true,
]{hyperref}

\usepackage[capitalize]{cleveref}

\newcommand{\CUSTOMvec}[1]{\mathbf{#1}}

\newcommand{\CUSTOMeps}{\epsilon}

\newcommand{\CUSTOMoneover}[1]{\frac{1}{#1}}

\newcommand{\CUSTOMtrm}[1]{\text{#1}}  \newcommand{\CUSTOMfor}{\quad \text{for}\;\;}    \newcommand{\CUSTOMwith}{\quad \text{with}\;\;} \newcommand{\CUSTOMra}{\rightarrow} \newcommand{\CUSTOMlimz}{\CUSTOMra 0} \newcommand{\CUSTOMlimi}{\CUSTOMra \infty}      

\newcommand{\CUSTOMhalf}{\frac{1}{2}}

\newcommand{\CUSTOMabs}[1]{\left\lvert #1 \right\rvert}   

\newcommand{\CUSTOMrbr}[1]{( #1 )}   \newcommand{\CUSTOMrbrl}[1]{\left( #1 \right)}

  \newcommand{\CUSTOMset}[1]{\{#1\}} 

\newcommand{\CUSTOMatval}[2]{\left.{#1}\right|_{#2}} 

\newcommand{\CUSTOMnn}[1]{\langle #1 \rangle}

   \newcommand{\CUSTOMopardiff}[1]{\frac{\partial}{\partial #1}}         

\newcommand{\CUSTOMdirac}{\delta}

 \newcommand{\CUSTOMT}{^{T}}      

  \newcommand{\CUSTOMpure}{\text{pure}}      \newcommand{\CUSTOMcrit}{c}

\newcommand{\CUSTOMexpect}[1]{\langle#1\rangle}    \newcommand{\CUSTOMcorr}[1]{\langle#1\rangle}

\newcommand{\CUSTOMerr}{\CUSTOMeps} \newcommand{\CUSTOMerrof}[1]{\CUSTOMerr{\CUSTOMrbr{#1}}} \newcommand{\CUSTOMmean}[1]{\overline{#1}} \newcommand{\CUSTOMmedian}{\CUSTOMtrm{md}} \newcommand{\CUSTOMmedianof}[1]{\CUSTOMmedian \CUSTOMrbr{#1}} 

\newcommand{\CUSTOMchisr}{\chi^2_{\CUSTOMtrm{red}}}

\newcommand{\CUSTOMnmeas}{N}    \newcommand{\CUSTOMncfg}{N_c}

    \newcommand{\CUSTOMprobdens}{p} \newcommand{\CUSTOMprobdensof}[1]{\CUSTOMprobdens\CUSTOMrbrl{#1}}

\newcommand{\CUSTOMavt}[1]{\langle#1\rangle} \newcommand{\CUSTOMavc}[1]{\left[#1\right]} \newcommand{\CUSTOMavtot}[1]{\CUSTOMavc{\CUSTOMavt{#1}}} \DeclareMathOperator{\CUSTOMrew}{\CUSTOMtrm{Rew}}

\newcommand{\CUSTOMO}{\mathcal{O}}    \newcommand{\CUSTOMham}{\mathcal{H}}

\newcommand{\CUSTOMdist}{r} \newcommand{\CUSTOMdistmin}{r_{\min}} \newcommand{\CUSTOMdistmax}{r_{\max}}  

\newcommand{\CUSTOMvol}{V} \newcommand{\CUSTOMoe}{e} \newcommand{\CUSTOMoE}{E} \newcommand{\CUSTOMoM}{M}  \newcommand{\CUSTOMom}{m} \newcommand{\CUSTOMoabsm}{\CUSTOMabs{\CUSTOMom}}       \newcommand{\CUSTOModufour}{\partial_{\beta}U_4} \newcommand{\CUSTOModutwo}{\partial_{\beta}U_2}  \newcommand{\CUSTOModlnm}{\partial_{\beta}(\ln \CUSTOMoabsm)}    \newcommand{\CUSTOMobetarew}{\Delta\beta_{\CUSTOMtrm{rew}}} \newcommand{\CUSTOMoratio}{Q}       \newcommand{\CUSTOMfitampl}{A}  \newcommand{\CUSTOMfitconst}{C} \newcommand{\CUSTOMfitfac}{A} \newcommand{\CUSTOMfitfactwo}{B}

   \newcommand{\CUSTOMcorrf}{C} \newcommand{\CUSTOMcorrdef}{\CUSTOMcorrf_{\CUSTOMdefect}}

 \newcommand{\CUSTOMbetac}{\beta_{c}} \newcommand{\CUSTOMbetasim}{\beta_{\CUSTOMtrm{sim}}} 

\newcommand{\CUSTOMkb}{{k_B}}  \newcommand{\CUSTOMikT}{ / (\CUSTOMkb T)}

\newcommand{\CUSTOMLmin}{L_{\min}}  

\newcommand{\CUSTOMpdef}{p_d} \newcommand{\CUSTOMpspin}{p_s} \newcommand{\CUSTOMpspinth}{\hat{p}_s} \newcommand{\CUSTOMpdefth}{\hat{p}_d} 

\newcommand{\CUSTOMpdefmean}{\CUSTOMmean{p}_d}  

\newcommand{\CUSTOMaexp}{a} \newcommand{\CUSTOMaexpmean}{\CUSTOMmean{\CUSTOMaexp}}  

\newcommand{\CUSTOMdim}{d} 

\newcommand{\CUSTOMspin}{s} \newcommand{\CUSTOMcoup}{J} \newcommand{\CUSTOMhfield}{h} \newcommand{\CUSTOMdefect}{\eta}

\newcommand{\CUSTOMeto}[1]{e^{#1}} 

  \newcommand{\CUSTOMpercent}[1]{\mbox{#1 \%}}

\newcommand{\CUSTOMeg}{\mbox{e.g.,}} 
\newcommand{\CUSTOMie}{\mbox{i.e.,}}

\newcommand{\CUSTOMdefhighlight}[1]{\emph{#1}}

\let\autocite\cite

\newcommand{\CUSTOMciteauthorref}[1]{\citeauthor{#1}~\autocite{#1}}
\newcommand{\CUSTOMciteauthorsref}[1]{\citeauthor{#1}~\autocite{#1}}

\newcommand{\CUSTOMciteauthorsyearref}[1]{\citeauthor{#1}~\citeyear{#1}~\autocite{#1}}

\begin{document}

\title{Critical exponent \texorpdfstring{$\nu$}{ν} of the Ising model in three dimensions with long-range correlated site disorder analyzed with Monte Carlo techniques}
\author{Stanislav Kazmin}
\email{kazmin@mis.mpg.de}
\affiliation{Max Planck Institute for Mathematics in the Sciences, Inselstrasse 22, 04103 Leipzig, Germany}
\affiliation{Institut für Theoretische Physik, Universität Leipzig, IPF 231101, 04081 Leipzig, Germany}
\author{Wolfhard Janke}
\affiliation{Institut für Theoretische Physik, Universität Leipzig, IPF 231101, 04081 Leipzig, Germany}
\date{\today}

\preprint{preprint number}

\begin{abstract}
	We study the critical behavior of the Ising model in three dimensions on a lattice with site disorder by using Monte Carlo simulations.
	The disorder is either uncorrelated or long-range correlated with correlation function that decays according to a power-law $\CUSTOMdist^{-\CUSTOMaexp}$.
	We derive the critical exponent of the correlation length $\nu$ and the confluent correction exponent $\omega$ in dependence of $\CUSTOMaexp$ by combining different  concentrations of defects $0.05 \leq \CUSTOMpdef \leq 0.4$ into one global fit ansatz and applying finite-size scaling techniques.
	We simulate and study a wide range of different correlation exponents $1.5 \leq \CUSTOMaexp \leq 3.5$ as well as the uncorrelated case $\CUSTOMaexp = \infty$ and are able to provide a global picture not yet known from previous works.
	Additionally, we perform a dedicated analysis of our long-range correlated disorder ensembles and provide estimates for the critical temperatures of the system in dependence of the correlation exponent $\CUSTOMaexp$ and the concentrations of defects $\CUSTOMpdef$.
	We compare our results to known results from other works and to the conjecture of Weinrib and Halperin: $\nu = 2/\CUSTOMaexp$ and discuss the occurring deviations.
\end{abstract}

\maketitle

\section{Introduction}

The influence of quenched disorder on phase transition properties of a system is of great importance as many real-world materials show defects or impurities.
The simplest way to introduce the disorder is by assuming it to be point-wise and uncorrelated.
A prominent achievement in describing the critical behavior of such systems was done by \CUSTOMciteauthorref{harris1974}.
The result is known as \CUSTOMdefhighlight{Harris criterion}.
It states that if the system has a negative specific heat exponent in the pure case (without disorder, $\alpha_{\CUSTOMpure} < 0$) the disorder does not influence the system's universality class.
On the other hand, for $\alpha_{\CUSTOMpure} > 0$ the disorder will change the system's universality class.
This universality class will have new critical exponents which will not depend on the disorder concentration.
Various studies \autocite{ballesteros1998,folk2003,calabrese2003,berche2004,murtazaev2004} confirmed the change of the universality class of the three-dimensional Ising model for which $\alpha_{\CUSTOMpure} > 0$ is true.

However, in nature the disorder usually comes with a certain structure.
One possible way to introduce such disorder to a model is by adding a spatial correlation to the disorder.
For a magnetic system this could be nonmagnetic lines or planes or clustered nonmagnetic impurities.
Other interesting areas are magnetic foams and magnetic elements in porous media \autocite{macfarland1996,paredes2006}.
The correlated disorder in systems was intensively studied with the help of the renormalization group theory by \CUSTOMciteauthorsref{weinrib1983} and the result is known as the \CUSTOMdefhighlight{extended Harris criterion}.
It states that a system with long-range correlated disorder where the spatial disorder correlation follows a power-law $\propto \CUSTOMdist^{-\CUSTOMaexp}$ will change its universality class if $\CUSTOMaexp < \CUSTOMdim$ and otherwise the standard Harris criterion will be recovered.
Further, they claim that the critical exponent of the correlation length $\nu$ in the long-range correlated three-dimensional Ising model is given by
\begin{align}
	\nu = \frac{2}{\CUSTOMaexp} \;.
	\label{eq:introduction:extended_harris_crit}
\end{align}
They argue, but do not prove rigorously, that this result is exact.
Several studies dealt with the Ising model with correlated disorder in two dimensions by applying Monte Carlo techniques \autocite{chatelain2014,chatelain2017} or renormalization group techniques \autocite{dudka2016}.
In three dimensions Monte Carlo simulations were performed in Refs.~\autocite{ballesteros1999,prudnikov2005,ivaneyko2008,herrmanns1999,ivaneyko2006,marques2009,wang2019} while renormalization group techniques were used in Refs.~\autocite{weinrib1983,prudnikov2000}.
While it is generally accepted that the correlated disorder case belongs to a new universality class, the quantitative results and in particular the claim given in \cref{eq:introduction:extended_harris_crit} are controversially discussed.
One condition which is often overseen when assuming \cref{eq:introduction:extended_harris_crit} is that $\CUSTOMdim = 4 - \epsilon \approx 4$ and $\CUSTOMaexp = 4 - \delta \approx 4$ is a necessary condition in Ref.~\autocite{weinrib1983}.
So it remains unclear which range of $\CUSTOMaexp$ values fulfills this requirement.
As a further reinforcement of the prediction given in \cref{eq:introduction:extended_harris_crit}, \CUSTOMciteauthorref{honkonen1989} claimed that \cref{eq:introduction:extended_harris_crit} is exact to all orders in the $\epsilon$-$\delta$-expansion.
This has been further analyzed in Refs.~\autocite{korzhenevskii1994,korzhenevskii1995}.

The results for the $\nu$ exponent obtained by different groups for the uncorrelated and the long-range correlated disordered three-dimensional Ising model are summarized in \cref{tab:introduction:disorder_nu_literature}.
The ambiguity about the numerical values of the critical exponents and considerable differences in the literature motivated us to attack the problem once again.

\begin{table*}
	\centering
	\caption[]{Various results of the critical exponent $\nu$ and the confluent correction exponent $\omega$ for the three-dimensional Ising model with uncorrelated and long-range power-law correlated disorder.
		We schematically denote the uncorrelated disorder case with $\CUSTOMaexp = \infty$.}
	\label{tab:introduction:disorder_nu_literature}
	\small
	\begin{tabular}{llllll}
	\toprule
	Reference                            & $\CUSTOMaexp$  & $\CUSTOMpdef$    & $\nu$      & $\omega$ & Remarks and Method                     \\
	\midrule
	\multicolumn{6}{c}{Uncorrelated Disorder}                                                                                     \\
	\midrule
	\CUSTOMciteauthorsyearref{ballesteros1998} & $\infty$ & 0.1 -- 0.4 & 0.6837(53) & 0.37(6)  &                                        \\
	\CUSTOMciteauthorsyearref{calabrese2003}   & $\infty$ & 0.2        & 0.690(8)   & $-$      &                                        \\
	\CUSTOMciteauthorsyearref{berche2004}      & $\infty$ & 0.3 -- 0.6 & 0.68(2)    & 0.7(1)   &                                        \\
	\CUSTOMciteauthorsyearref{murtazaev2004}   & $\infty$ & 0.2        & 0.683(4)   & $-$      & also values for $\CUSTOMpdef$ = 0.1 -- 0.4   \\
	\midrule
	\multicolumn{6}{c}{Correlated Disorder}                                                                                       \\
	\midrule
	\CUSTOMciteauthorsyearref{weinrib1983}     & 2.0      & $-$        & 1          & $-$      & one-loop $\epsilon$-$\delta$-expansion \\
	\CUSTOMciteauthorsyearref{prudnikov2000}   & 2.0      & 0.2        & 0.7151     & $-$      & two-loop massive renormalization       \\
	\midrule
	\CUSTOMciteauthorsyearref{ballesteros1999} & 2.0      & 0.2, 0.35  & 1.012(10)  & 1.01(13) & point-wise power-law                   \\
	\CUSTOMciteauthorsyearref{prudnikov2005}   & 2.0      & 0.2        & 0.710(10)  & 0.8      & defect lines                           \\
	\CUSTOMciteauthorsyearref{ivaneyko2008}    & 2.0      & 0.2        & 0.958(4)   & 0.8      & defect lines / point-wise power-law    \\
	\bottomrule
\end{tabular} \end{table*}

We extensively analyzed a three-dimensional Ising lattice with power-law correlated site disorder by using Monte Carlo techniques.
In contrast to previous works we performed simulations for various different correlation strengths $\CUSTOMaexp$ and a wide range of disorder concentrations $\CUSTOMpdef$.
We focused on the critical exponent of the correlation length $\nu$ and the confluent correction exponent $\omega$ and obtained a global picture of their behaviors in the long-range correlated cases and in the uncorrelated disorder case.
Additionally, we can present a rich palette of critical temperatures for various $\CUSTOMaexp$ and $\CUSTOMpdef$.

The rest of the paper is structured as follows.
In Section~\ref{sec:model_and_simulation} we specify our model and the details of the performed simulations.
In Section~\ref{sec:cfg_analysis} we analyze the disorder realizations to confirm the desired power-law behavior.
The main analysis of the Monte Carlo simulations of the Ising model and the obtained results are contained in Section~\ref{sec:critical_expoent_nu}.
We present the derivation of the critical exponent $\nu$ as well as the correction exponent $\omega$.
We compare our results to the Weinrib and Halperin conjecture, $\nu=2/\CUSTOMaexp$, and to the known results.
Finally we obtain critical temperatures for different concentrations and correlation exponents.
A conclusion in Section~\ref{sec:conclusions} completes this work.

{
\newcommand{\CUSTOMx}{x}
\newcommand{\CUSTOMy}{y}

\section{Model and Simulation Details}
\label{sec:model_and_simulation}

\subsection{Ising Model with Site Disorder}

We will not discuss the standard Ising model here and refer to \autocite{landau2009,newman1999a} as a good starting point for readers who need a deeper background.
For the rest of the paper we will deal with the Ising model with site disorder which we will refer to as \CUSTOMdefhighlight{disordered Ising model}\footnote{
	The Ising model with random couplings, \CUSTOMie{} bond disorder, is also called \enquote{disordered Ising model} in the literature.}.
The Hamiltonian of the Ising model with site disorder has a very similar form to the standard Ising model
\begin{align}
	\CUSTOMham = - \CUSTOMcoup \sum_{\CUSTOMnn{\CUSTOMx\CUSTOMy}} \CUSTOMdefect_\CUSTOMx \CUSTOMdefect_\CUSTOMy \CUSTOMspin_\CUSTOMx \CUSTOMspin_\CUSTOMy - \CUSTOMhfield \sum_\CUSTOMx \CUSTOMdefect_\CUSTOMx \CUSTOMspin_\CUSTOMx\;,
	\label{eq:model_and_simulation:disordered_ising_model_hamiltonian_def}
\end{align}
where the spins can take the values $\CUSTOMspin_{\CUSTOMx} = \pm1$ and the defect variables can be $\CUSTOMdefect_{\CUSTOMx} = 1$ when the spin is present at site $\CUSTOMx$ and $\CUSTOMdefect_{\CUSTOMx} = 0$ when the site $\CUSTOMx$ is empty (a defect).
The sum runs over all next-neighbors denoted by $\CUSTOMnn{\CUSTOMx\CUSTOMy}$.
The coupling constant is set to $\CUSTOMcoup = 1$ on the whole lattice and we work without an external magnetic field, \CUSTOMie{} \mbox{$\CUSTOMhfield = 0$}.
Schematically the Ising model with and without site disorder is presented in \cref{fig:model_and_simulation:disordered_ising_model_scheme}.

\begin{figure}[t]
\centering
	\subfloat[$\CUSTOMpdef = 0$.]{\includegraphics[scale=1]{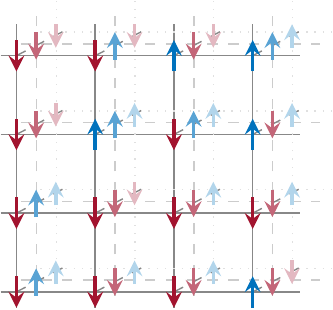}
 \label{fig:model_and_simulation:disordered_ising_model_scheme_a}}
	\subfloat[$\CUSTOMpdef \approx 0.35$.]{\includegraphics[scale=1]{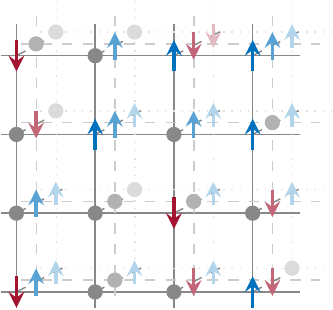}
 \label{fig:model_and_simulation:disordered_ising_model_scheme_b}}

	\caption[]{Three-dimensional Ising model lattices \subref{fig:model_and_simulation:disordered_ising_model_scheme_a} without and \subref{fig:model_and_simulation:disordered_ising_model_scheme_b} with site disorder.
		The red and blue arrows represent the spins with the states $\CUSTOMspin_\CUSTOMx = \pm 1$, respectively.
		The gray points represent the defects (vacant sites).}
	\label{fig:model_and_simulation:disordered_ising_model_scheme}
\end{figure}

We distinguish between two different disorder types.
The first type is the \CUSTOMdefhighlight{uncorrelated disorder} or random disorder.
In this case the defects are chosen randomly according to the probability density
\begin{align}
	\CUSTOMprobdensof{\CUSTOMdefect} = \CUSTOMpspin \CUSTOMdirac(\CUSTOMdefect) + \CUSTOMpdef \CUSTOMdirac(\CUSTOMdefect - 1) \;,
	\label{eq:model_and_simulation:uncorr_def_distribution}
\end{align}
where $\CUSTOMpspin$ is the concentration of spins, $\CUSTOMpdef = 1 - \CUSTOMpspin$ is the concentration of defects and $\CUSTOMdirac$ is the Dirac-delta distribution.

The second type is the \CUSTOMdefhighlight{correlated disorder}.
In this case the probability density for the defects is again given by \cref{eq:model_and_simulation:uncorr_def_distribution}.
However, now additionally the spatial correlation between the defects decays according to a power-law
\begin{align}
	\CUSTOMexpect{\CUSTOMdefect_\CUSTOMx\CUSTOMdefect_\CUSTOMy} \propto \CUSTOMoneover{\CUSTOMdist(\CUSTOMx,\CUSTOMy)^\CUSTOMaexp} \;,
	\label{eq:model_and_simulation:corr_powerlaw_def}
\end{align}
where $\CUSTOMdist(\CUSTOMx,\CUSTOMy)$ is the distance between sites $\CUSTOMx$ and $\CUSTOMy$ and $\CUSTOMaexp \geq 0$ is the \CUSTOMdefhighlight{correlation exponent}.
Note, that for both cases we work in the so-called grand-canonical approach where the desired concentration $\CUSTOMpdef$ is a mean value over a large number of realizations while for each separate realization $\CUSTOMpdef$ can vary.
In \cref{fig:model_and_simulation:corr_defect_slices} we show slices of a three-dimensional Ising model lattice with different concentrations of defects and different correlation exponents near the critical temperature.

\begin{figure}[t]
	\captionsetup[subfigure]{labelformat=empty}
	\centering
	\subfloat[$\CUSTOMaexp = \infty$, $\CUSTOMpdef = 0$.]{\includegraphics[scale=1]{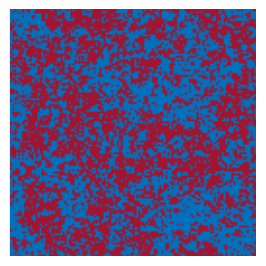}}
	\hfill
	\subfloat[$\CUSTOMaexp = \infty$, $\CUSTOMpdef = 0.2$.]{\includegraphics[scale=1]{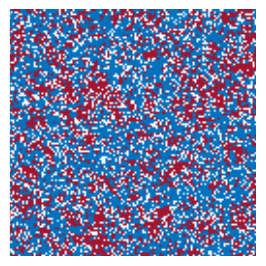}}
	\hfill
	\subfloat[$\CUSTOMaexp = \infty$, $\CUSTOMpdef = 0.4$.]{\includegraphics[scale=1]{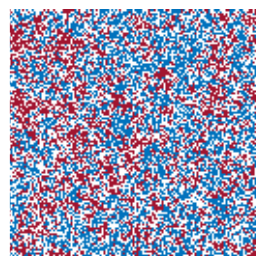}}

	\subfloat[]{\hspace*{2.7cm}}
	\hfill
	\subfloat[$\CUSTOMaexp = 1.5$, $\CUSTOMpdef = 0.2$.]{\includegraphics[scale=1]{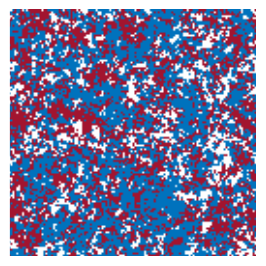}}
	\hfill
	\subfloat[$\CUSTOMaexp = 1.5$, $\CUSTOMpdef = 0.4$.]{\includegraphics[scale=1]{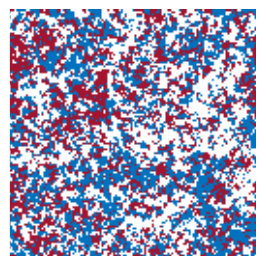}}

	\caption[]{Slices of a three-dimensional Ising model lattice with $L=128$ simulated near the critical temperature for different correlation exponents $\CUSTOMaexp$ and concentrations of defects $\CUSTOMpdef$.
		Red and blue points represent the spin states $\CUSTOMspin_\CUSTOMx = \pm1$ and white points represent the defects $\CUSTOMdefect_\CUSTOMx = 0$.
		One can see that correlated defects tend to form clusters of defects.}
	\label{fig:model_and_simulation:corr_defect_slices}
\end{figure}

According to the Harris criterion and the extended Harris criterion the disordered three-dimensional Ising model falls into three different universality classes in dependence of the correlation exponent $\CUSTOMaexp$ and the concentration of defects $\CUSTOMpdef$.
The pure case where no defects are present ($\CUSTOMpdef = 0$), the effectively uncorrelated case for $\CUSTOMaexp > \CUSTOMdim$ and the correlated case for $\CUSTOMaexp \leq \CUSTOMdim$.
These cases are schematically shown in \cref{fig:model_and_simulation:a_pd_diagram}.

\begin{figure}[t]
	\centering
	\includegraphics[scale=1]{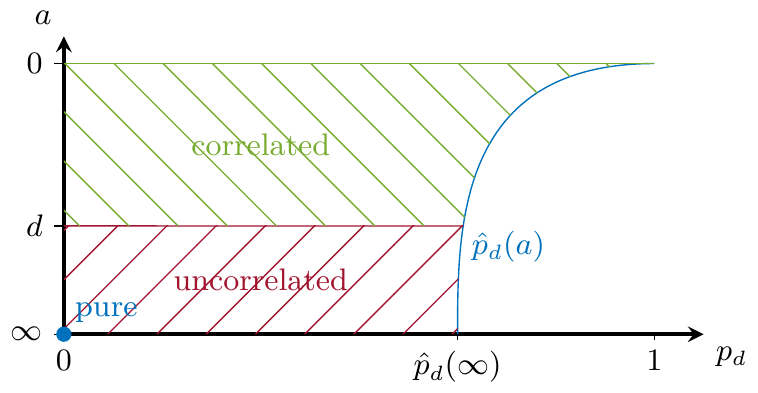}
 	\caption[]{Universality classes of the three-dimensional Ising model for different correlation exponents $\CUSTOMaexp$ and concentrations of defects $\CUSTOMpdef$.
		The curve $\CUSTOMpdefth(\CUSTOMaexp) = 1- \CUSTOMpspinth(\CUSTOMaexp)$ is the percolation threshold of the defect concentration below which an infinite spin cluster exists for $L \CUSTOMlimi$.
		It has been shown in Ref.~\autocite{zierenberg2017} that for smaller $\CUSTOMaexp$ values the concentration of spins $\CUSTOMpspin$ can be chosen lower without destroying the infinite cluster, thus $\CUSTOMpdef$ increases for stronger correlations (smaller $\CUSTOMaexp$).}
	\label{fig:model_and_simulation:a_pd_diagram}
\end{figure}

\subsection{Monte Carlo Simulation Details}

We performed Monte Carlo simulations of the disordered Ising model and used the Swendsen-Wang multi-cluster update algorithm \autocite{swendsen1987}.
The linear lattice sizes of our cubic lattices were in the range between $L = 8$ and $L = 256$ and we chose periodic boundary conditions in each direction.
The correlation exponent values were $\CUSTOMaexp = $~1.5, 2.0, 2.5, 3.0, 3.5 and $\infty$ which we will use symbolically for the uncorrelated case.
For each $\CUSTOMaexp$ value we simulated eight defect concentrations $\CUSTOMpdef = $~0.05, 0.1, 0.15, 0.2, 0.25, 0.3, 0.35 and 0.4.
After a thermalization period of \num{500} sweeps we performed $\CUSTOMnmeas = \num{10000}$ measurement sweeps at each considered simulation point $\CUSTOMbetasim = 1 / (\CUSTOMkb T_{\CUSTOMtrm{sim}})$.
Throughout the paper we will refer to the inverse temperature defined by $\beta = 1 / (\CUSTOMkb T)$ simply as \enquote{temperature}.
The temperatures were first chosen in a wide range and with larger spacing for small lattices.
After the first analyses refined ranges (regions around the critical points for considered observables for finite lattice sizes $L$) were estimated and larger lattice sizes were simulated at less temperatures.
For each parameter tuple $(\CUSTOMaexp, \CUSTOMpdef, L)$ we simulated $\CUSTOMncfg = \num{1000}$ disorder realizations.
After each sweep we measured and stored the total energy $\CUSTOMoE$
\begin{align}
	\CUSTOMoE = - \CUSTOMcoup \sum_{\CUSTOMnn{\CUSTOMx\CUSTOMy}} \CUSTOMdefect_\CUSTOMx \CUSTOMdefect_\CUSTOMy \CUSTOMspin_\CUSTOMx \CUSTOMspin_\CUSTOMy \;,
	\label{eq:model_and_simulation:E_def}
\end{align}
and the total magnetization of the system $\CUSTOMoM$
\begin{align}
	\CUSTOMoM = \sum_{\CUSTOMx} \CUSTOMdefect_\CUSTOMx \CUSTOMspin_\CUSTOMx  \;.
	\label{eq:model_and_simulation:M_def}
\end{align}
At the end we had two-dimensional arrays of values $\CUSTOMoE_i^c$ and $\CUSTOMoM_i^c$ where $i = 1, \dots, \CUSTOMnmeas$ and $c = 1, \dots, \CUSTOMncfg$ for each parameter tuple $(\CUSTOMaexp, \CUSTOMpdef, L, \CUSTOMbetasim)$.
This was needed in order to apply the reweighting technique in later analysis.

}

\section{Correlated Disorder Analysis}
\label{sec:cfg_analysis}

Before we move to the analysis of the Monte Carlo simulations of the Ising model we first take a look at the site disorder realization and analyze the generated ensembles.
It is a necessary step to gain control over the correlation exponents $\CUSTOMaexp$ of the disorder ensembles on which we will perform the simulations later on.

\subsection{Disorder Generation}

In this work we mainly study the Ising model on a lattice with uncorrelated and long-range correlated site disorder.
An important part is the generation of the site disorder for later Monte Carlo simulations.
The \CUSTOMdefhighlight{uncorrelated disorder} case is realized by setting the defect variables $\CUSTOMdefect_x$ for each site $x$ of the lattice according to
\begin{align}
	\CUSTOMdefect_x = \begin{cases}
		0 \quad \CUSTOMtrm{if } R_x \leq \CUSTOMpdef \\
		1 \quad \CUSTOMtrm{else}
	\end{cases} \;,
	\label{eq:cfg_analysis:defect_values_def}
\end{align}
where $0 \leq R_x < 1$ is a uniform random number drawn for each site $x$.

For the case of \CUSTOMdefhighlight{long-range correlated disorder} let us first define the correlation function $\CUSTOMcorrdef$ between two defects $\CUSTOMdefect$ at sites $x$ and $y$ at a distance $\CUSTOMdist = \CUSTOMabs{x-y}$
\begin{align}
	\CUSTOMcorrdef(\CUSTOMdist) = \CUSTOMatval{\CUSTOMcorr{\CUSTOMdefect(x) \CUSTOMdefect(y)}}{\CUSTOMabs{x-y}=\CUSTOMdist} \;.
	\label{eq:cfg_analysis:general_corrf_disorder}
\end{align}
In this work we assume a \CUSTOMdefhighlight{power-law decay} of the correlation function for large distances $\CUSTOMdist \gg 1$
\begin{align}
	\CUSTOMcorrdef(\CUSTOMdist) \propto \CUSTOMdist^{-\CUSTOMaexp} \;.
	\label{eq:cfg_analysis:power_law_decay_corrf_disorder}
\end{align}
We used a modified Fourier method by \CUSTOMciteauthorref{zierenberg2017} for the generation of long-range correlated disorder.
Initially the Fourier method was introduced by \CUSTOMciteauthorref{makse1995}.
The code linked in \autocite{zierenberg2017} was used in this work.
We will not discuss the details of the generation and only sketch the process:
\begin{enumerate}
	\item Generate uncorrelated, normally distributed random variables.
	\item Perform a Fourier transformation of these variables.
	\item Correlate the transformed variables by multiplying with a spectral density generated from a chosen correlation function $\CUSTOMcorrf_0$.
	\item Fourier transform the correlated variables back to real space.
	\item Truncate the final variables to $\CUSTOMset{0,1}$ such that the mean concentration of zeros equals the desired concentration of defects $\CUSTOMpdef$.
\end{enumerate}

The resulting $\CUSTOMdefect$ variables are correlated and their correlation function is approximately given by $\CUSTOMcorrf_0$.
This approximation comes from the fact that the truncation in step 5 is not mathematically exact and introduces deviations from the desired function $\CUSTOMcorrf_0$.

In order to overcome the infinity at $\CUSTOMdist = 0$ in \cref{eq:cfg_analysis:power_law_decay_corrf_disorder} we used a slightly modified correlation function
\begin{align}
	\CUSTOMcorrf_0(\CUSTOMdist) \propto \CUSTOMrbrl{1 + \CUSTOMdist^2}^{-\CUSTOMaexp/2} \;,
	\label{eq:cfg_analysis:modified_power_law_corrf_disorder}
\end{align}
which asymptotically approaches \cref{eq:cfg_analysis:power_law_decay_corrf_disorder} for large distances,
\begin{align}
	\CUSTOMcorrf_0(\CUSTOMdist) \CUSTOMra \CUSTOMdist^{-\CUSTOMaexp} \CUSTOMfor \CUSTOMdist \CUSTOMlimi \;.
	\label{eq:cfg_analysis:modified_power_law_corrf_limit}
\end{align}

We generated ensembles of disorder realizations by providing two parameters: the correlation decay exponent $\CUSTOMaexp$ and the concentration of defects $\CUSTOMpdef$.
Because the step 5 is mathematically not exact and also because a modified correlation function was used, we verified   both values $\CUSTOMaexp$ and $\CUSTOMpdef$ for each ensemble numerically.

\subsection{Mean Concentration of Defects}

\begin{figure}[t]
	\centering
	\subfloat[$\CUSTOMaexp = 2.0$, $\CUSTOMpdef = 0.2$, $L = 8$.]{\includegraphics[scale=1]{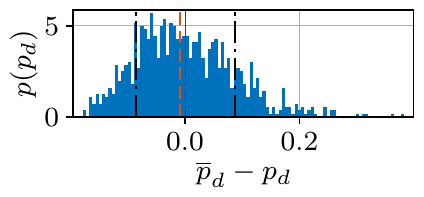}}
	\subfloat[$\CUSTOMaexp = \infty$, $\CUSTOMpdef = 0.2$, $L = 8$.]{\includegraphics[scale=1]{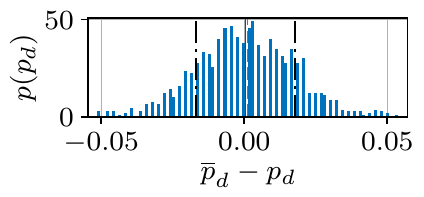}}

	\subfloat[$\CUSTOMaexp = 2.0$, $\CUSTOMpdef = 0.2$, $L = 32$.]{\includegraphics[scale=1]{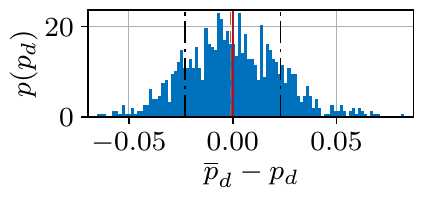}}
	\subfloat[$\CUSTOMaexp = \infty$, $\CUSTOMpdef = 0.2$, $L = 32$.]{\includegraphics[scale=1]{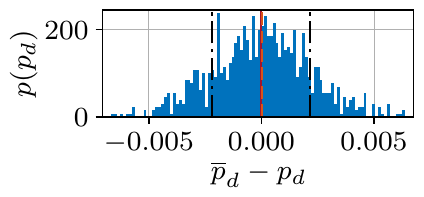}}

	\subfloat[$\CUSTOMaexp = 2.0$, $\CUSTOMpdef = 0.2$, $L = 256$.]{\includegraphics[scale=1]{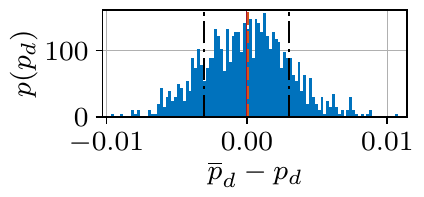}}
	\subfloat[$\CUSTOMaexp = \infty$, $\CUSTOMpdef = 0.2$, $L = 256$.]{\includegraphics[scale=1]{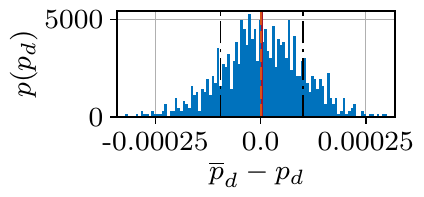}}

	\subfloat{\includegraphics[scale=1]{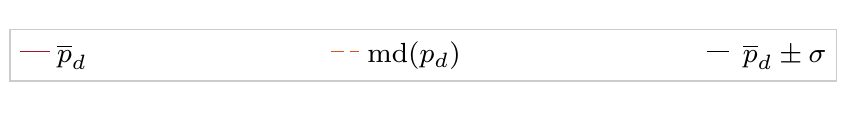}}

	\caption[]{Histograms of concentrations of defects $\CUSTOMpdef$ for different parameter tuples.
		$\CUSTOMmedianof{\CUSTOMpdef}$ is the median of the ensemble and $\CUSTOMpdef$ is the imposed concentration value whereas $\CUSTOMpdefmean$ is the calculated mean.}
	\label{fig:cfg_analysis:pdef_distribution}
\end{figure}

\begin{figure}[t]
	\centering
	\subfloat[$\CUSTOMaexp = \infty$.]{\includegraphics[scale=1,valign=t]{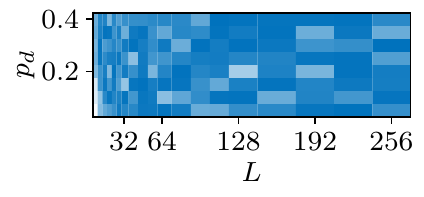}}
	\subfloat[$\CUSTOMaexp = 3.5$.]{\includegraphics[scale=1,valign=t]{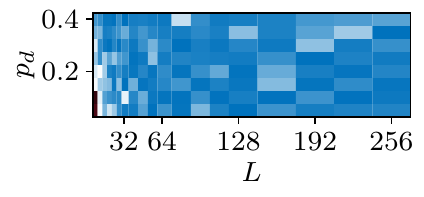}}

	\subfloat[$\CUSTOMaexp = 3.0$.]{\includegraphics[scale=1]{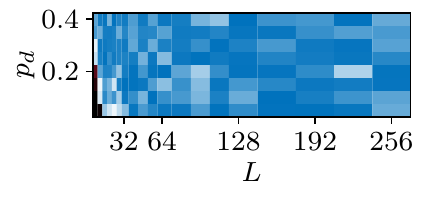}}
	\subfloat[$\CUSTOMaexp = 2.5$.]{\includegraphics[scale=1]{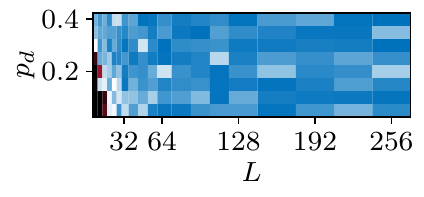}}

	\subfloat[$\CUSTOMaexp = 2.0$.]{\includegraphics[scale=1]{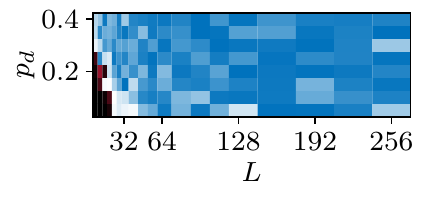}}
	\subfloat[$\CUSTOMaexp = 1.5$.]{\includegraphics[scale=1]{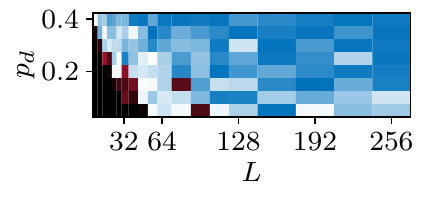}}

	\subfloat{\includegraphics[scale=1,valign=t]{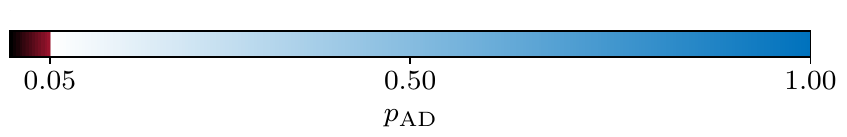}}

	\caption[]{Anderson-Darling test results for different parameter tuples ($\CUSTOMaexp$, $\CUSTOMpdef$, $L$).
		$p_{\CUSTOMtrm{AD}}$ is the probability according to Anderson-Darling test for the analyzed variables to come from a normal distribution.
		Black and red regions suggest non-normal distributions while white and blue regions suggest normal distributions.}
	\label{fig:cfg_analysis:pdf_distribution_test}
\end{figure}

First we looked at the distribution of the concentrations of defects $\CUSTOMpdef$ for each parameter tuple ($\CUSTOMaexp$, $\CUSTOMpdef$, $L$).
Examples of the distributions are shown in \cref{fig:cfg_analysis:pdef_distribution}.
We verified the normality of the distributions for each ensemble with the help of the Anderson-Darling test \autocite{anderson1952,thode2002}.
Apart from the strongest correlation with $\CUSTOMaexp = 1.5$ at low $\CUSTOMpdef \leq 0.2$ all distributions for $L \geq 24$ were classified as normal with 95~\% confidence.
The results of the test for all parameter tuples ($\CUSTOMaexp$, $\CUSTOMpdef$, $L$) are presented in \cref{fig:cfg_analysis:pdf_distribution_test}.
It can be seen that higher concentrations approach the normal distribution already for smaller $L$.
The estimated concentrations $\CUSTOMpdefmean$ as a mean over all lattice sizes for each ensemble are listed in \cref{tab:cfg_analysis:pd_mean_final}.
They match the imposed concentrations $\CUSTOMpdef$ perfectly in all cases.

\begin{table}[t]
	\centering
	\small
	\caption[]{Summary of obtained mean concentrations $\CUSTOMpdefmean$ for all $\CUSTOMaexp$ and $\CUSTOMpdef$ parameters.
		The means were calculated over all lattice sizes with $L \geq 24$.}
	\label{tab:cfg_analysis:pd_mean_final}
	\sisetup{
	table-text-alignment=center,
	table-format = 1.7,
}
$\begin{array}{lSSS}
\toprule
 \CUSTOMpdef & {\CUSTOMaexp = \infty} & {\CUSTOMaexp = 3.5} & {\CUSTOMaexp = 3.0} \\
\midrule
 0.05  & 0.05001(3)       & 0.05001(7)    & 0.05002(8)    \\
 0.1   & 0.10000(4)       & 0.1000(1)     & 0.1000(2)     \\
 0.15  & 0.15001(6)       & 0.1500(1)     & 0.1501(2)     \\
 0.2   & 0.20000(4)       & 0.2000(2)     & 0.2000(3)     \\
 0.25  & 0.24999(5)       & 0.2500(2)     & 0.2501(4)     \\
 0.3   & 0.29999(8)       & 0.3001(3)     & 0.3000(2)     \\
 0.35  & 0.35001(9)       & 0.3500(3)     & 0.3501(4)     \\
 0.4   & 0.4000(1)        & 0.4001(4)     & 0.3999(4)     \\
\midrule
 \CUSTOMpdef & {\CUSTOMaexp = 2.5}    & {\CUSTOMaexp = 2.0} & {\CUSTOMaexp = 1.5} \\
\midrule
 0.05  & 0.0500(2)        & 0.0500(3)     & 0.0499(5)     \\
 0.1   & 0.1000(3)        & 0.1001(4)     & 0.1000(8)     \\
 0.15  & 0.1500(3)        & 0.1499(6)     & 0.1498(9)     \\
 0.2   & 0.1999(3)        & 0.2002(7)     & 0.200(2)      \\
 0.25  & 0.2500(4)        & 0.2499(6)     & 0.250(2)      \\
 0.3   & 0.2999(5)        & 0.3001(7)     & 0.300(2)      \\
 0.35  & 0.3501(6)        & 0.3501(7)     & 0.350(2)      \\
 0.4   & 0.4001(6)        & 0.400(2)      & 0.401(2)      \\
\bottomrule
\end{array}$ \end{table}

\subsection{Mean Correlation Exponent}

The correlation function $\CUSTOMcorrdef(\CUSTOMdist)$ was calculated as a mean over all configurations for each parameter tuple ($\CUSTOMaexp$, $\CUSTOMpdef$, $L$).
It was measured for two different distance directions (along the $x$-axis and along the diagonal),
\begin{align}
	\hat{\CUSTOMvec{\CUSTOMdist}}_1 & = (1, 0, 0)\CUSTOMT \;, \quad \hat{\CUSTOMvec{\CUSTOMdist}}_2  = (1, 1, 1)\CUSTOMT \;,
	\label{eq:cfg_analysis:corr_func_dist_dir_vectors}
\end{align}
and all possible distances in the corresponding direction.
The correlation function was calculated by
\begin{align}
	\CUSTOMcorrdef(\CUSTOMdist)
	= \CUSTOMexpect{\CUSTOMdefect_x \CUSTOMdefect_y}
	= \left<
	\frac{\mathcal{C} }{ N_{\CUSTOMdist}} \sum_{\substack{x,y \\ y - x = \CUSTOMdist \hat{\CUSTOMvec{\CUSTOMdist}}_i}} (\CUSTOMdefect_x - \CUSTOMpdefmean)(\CUSTOMdefect_y - \CUSTOMpdefmean)
	\right> \;,
	\label{eq:cfg_analysis:corr_func_for_defects_definition}
\end{align}
where $\mathcal{C}$ is the normalization constant such that $\CUSTOMcorrdef(0) = 1$ and $N_{\CUSTOMdist}$ is the number of possible realizations of the distance $\CUSTOMdist$ on the lattice.
From the chosen $\hat{\CUSTOMvec{\CUSTOMdist}}_i$ vectors and from periodic boundary conditions it follows that
\begin{align}
	N_{\CUSTOMdist} = \begin{cases}
		\CUSTOMvol / 2 & \CUSTOMtrm{for } \CUSTOMdist = L / 2 \;\CUSTOMtrm{and}\; \CUSTOMdist = \sqrt{3} L / 2 \\
		\CUSTOMvol     & \CUSTOMtrm{else}
	\end{cases}\;.
	\label{eq:cfg_analysis:number_of_dist_realizations}
\end{align}
The sum in \cref{eq:cfg_analysis:corr_func_for_defects_definition} runs over all site pairs $x$ and $y$ which have the vector distance $\CUSTOMdist \hat{\CUSTOMvec{\CUSTOMdist}}_i$ where $i = 1,2$.
The normalization constant turns out to be
\begin{align}
	\mathcal{C} =  \CUSTOMoneover{\CUSTOMpdefmean ( 1 - \CUSTOMpdefmean)} \;.
	\label{eq:cfg_analysis:corr_function_normalization_const}
\end{align}

Once the correlation functions defined through \cref{eq:cfg_analysis:corr_func_for_defects_definition} were measured for each disorder ensemble, we had to obtain the correlation exponent $\CUSTOMaexp$.
We performed a fit to the linearized ansatz on a logarithmic scale corresponding to the asymptotic behavior of \cref{eq:cfg_analysis:power_law_decay_corrf_disorder}
\begin{align}
	\ln \CUSTOMcorrdef (\CUSTOMdist) = - \CUSTOMaexp \ln \CUSTOMdist + \CUSTOMfitfactwo \;,
	\label{eq:cfg_analysis:loglog_corr_func_fit_ansatz}
\end{align}
where $\CUSTOMaexp$ is the desired decay exponent.
We had to find a minimal distance $\CUSTOMdistmin$ included into the fits in order to obtain the correlation exponent for $\CUSTOMdist > \CUSTOMdistmin \gg 1$ where the assumption of a power-law decay is valid.
We used the condition that $\CUSTOMdistmin$ is the distance where the relative deviation between $\CUSTOMcorrf_0$ and $\CUSTOMcorrdef$ became less than a threshold value of $\CUSTOMerr_{\CUSTOMcorrf} = 0.05$ for the first time,
\begin{align}
	  & \frac{\CUSTOMcorrdef(\CUSTOMdistmin) - \CUSTOMcorrf_0(\CUSTOMdistmin)}{\CUSTOMcorrdef(\CUSTOMdistmin)} \notag                                 \\
	= & \frac{\CUSTOMdistmin^{-\CUSTOMaexp} - (1 + \CUSTOMdistmin^2)^{-\CUSTOMaexp / 2}}{\CUSTOMdistmin^{-\CUSTOMaexp}} \leq \CUSTOMerr_{\CUSTOMcorrf} = 0.05 \;.
	\label{eq:cfg_analysis:corr_func_rel_error_rmin}
\end{align}
Note that the amplitudes for $\CUSTOMcorrf_0$ and $\CUSTOMcorrdef$ were omitted as we assume them to be equal and to cancel in \cref{eq:cfg_analysis:corr_func_rel_error_rmin}.
\cref{eq:cfg_analysis:corr_func_rel_error_rmin} leads to the condition
\begin{align}
	\CUSTOMdistmin(\CUSTOMaexp) \geq \CUSTOMrbrl{(1 - \CUSTOMerr_{\CUSTOMcorrf})^{-2/\CUSTOMaexp} - 1}^{-1/2} \;.
	\label{eq:cfg_analysis:rmin_from_rel_err}
\end{align}

Furthermore we had to set a maximum distance $\CUSTOMdistmax$ to exclude the noisy tail of the correlation function and possible finite-size effects.
Here we have chosen the distance $\CUSTOMdistmax$ where the absolute value of the measured correlation function $\CUSTOMabs{\CUSTOMcorrdef}$ was below a minimal threshold value of $\CUSTOMcorrf_{\min} = 10^{-5}$ for the first time,
\begin{align}
	\CUSTOMabs{\CUSTOMcorrdef\CUSTOMrbrl{\CUSTOMdistmax}} \leq \CUSTOMcorrf_{\min} = 10^{-5}\;.
	\label{eq:cfg_analysis:rmax_condition_def}
\end{align}

\begin{figure}[t]
	\centering
	\subfloat[$\CUSTOMaexp = 3.5$.]{\includegraphics[scale=1]{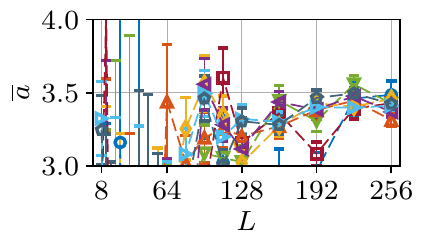}}
	\subfloat[$\CUSTOMaexp = 3.0$.]{\includegraphics[scale=1]{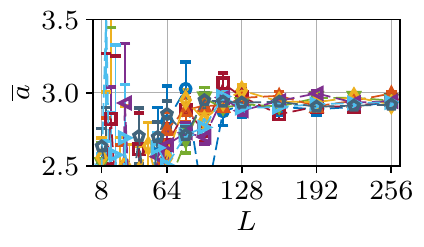}}

	\subfloat[$\CUSTOMaexp = 2.5$.]{\includegraphics[scale=1]{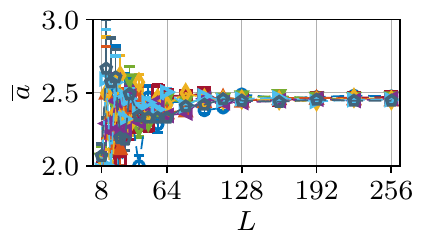}}
	\subfloat[$\CUSTOMaexp = 2.0$.]{\includegraphics[scale=1]{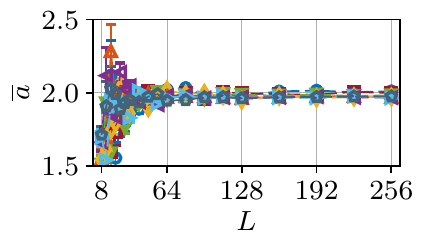}}

	\subfloat[$\CUSTOMaexp = 1.5$.]{\includegraphics[scale=1]{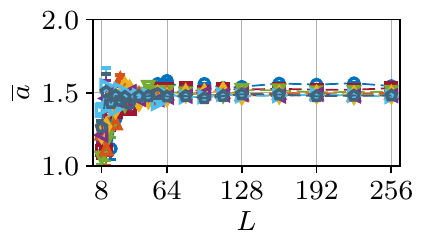}}
	\subfloat{\includegraphics[scale=1]{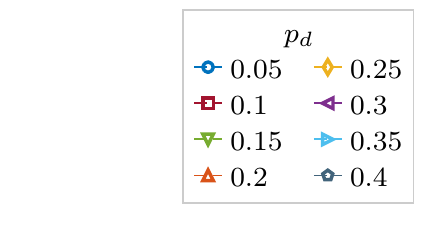}}

	\caption[]{Measured correlation exponents $\CUSTOMaexpmean$ for different concentrations of defects $\CUSTOMpdef$ and chosen correlation exponents $\CUSTOMaexp$.
		Larger lattices have more possibilities to realize a certain distance $\CUSTOMdist$ and therefore the estimates $\CUSTOMaexpmean(L)$ become statistically better with increasing $L$.
		For weak correlations (large $\CUSTOMaexp$) only the largest lattices $L \gtrapprox 160$ approach the expected values $\CUSTOMaexp$.
		Dashed lines are shown to guide the eye.}
	\label{fig:cfg_analysis:amean_estimates}
\end{figure}

\begin{figure}[t]
	\centering
	\subfloat[$\CUSTOMaexp = 2.0$, $\CUSTOMpdef = 0.2$, $L = 32$.]{\includegraphics[scale=1]{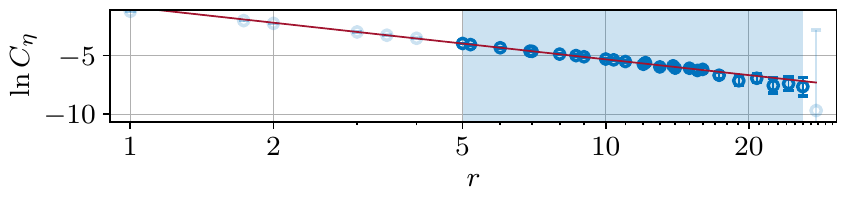}}

	\subfloat[$\CUSTOMaexp = 2.0$, $\CUSTOMpdef = 0.2$, $L = 256$.]{\includegraphics[scale=1]{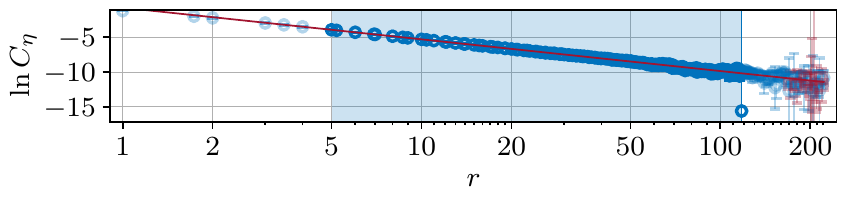}}

	\subfloat[$\CUSTOMaexp = 3.5$, $\CUSTOMpdef = 0.2$, $L = 32$.]{\includegraphics[scale=1]{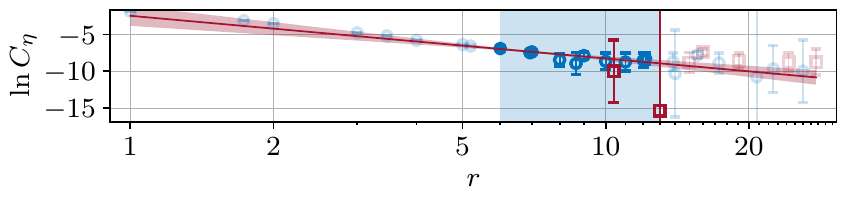}}

	\subfloat[$\CUSTOMaexp = 3.5$, $\CUSTOMpdef = 0.2$, $L = 256$.]{\includegraphics[scale=1]{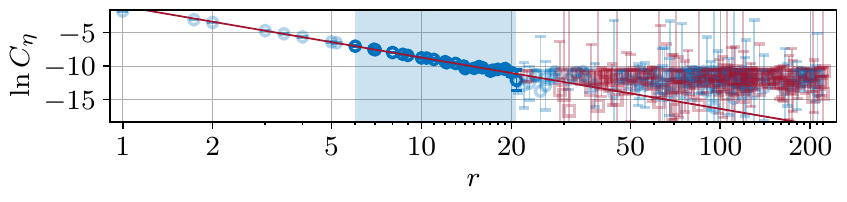}}

	\subfloat{\includegraphics[scale=1]{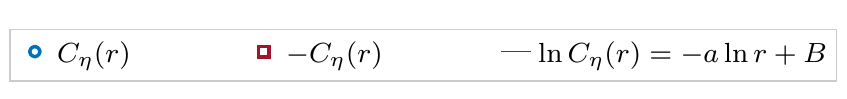}}

	\caption[]{Fits of the correlation of defects to the ansatz  $\ln \CUSTOMcorrdef (\CUSTOMdist) = - \CUSTOMaexp \ln \CUSTOMdist + \CUSTOMfitfactwo$, \cref{eq:cfg_analysis:loglog_corr_func_fit_ansatz}, for different parameters.
		Fits to weaker correlations (larger $\CUSTOMaexp$) use less points because the signal gets noisy faster.
		This leads to larger errors compared to lower $\CUSTOMaexp$ values.
		The blue regions show the regions between $\CUSTOMdistmin$ and $\CUSTOMdistmax$.
		The maximum distance on the x-axis is the distance along the diagonal with $\CUSTOMdist = \sqrt{3} L / 2$.
	}
	\label{fig:cfg_analysis:corr_func_fit}
\end{figure}

For small lattices with $L \leq 20$ and weak correlations (large $\CUSTOMaexp$) sometimes the found $\CUSTOMdistmin$ and $\CUSTOMdistmax$ where too close together or even $\CUSTOMdistmin > \CUSTOMdistmax$.
Is such cases we reduced $\CUSTOMdistmin$ until a fit with 4 degrees of freedom was possible.
The estimated $\CUSTOMaexpmean(\CUSTOMpdef,L)$ are shown in \cref{fig:cfg_analysis:amean_estimates} and the final averages are summarized in \cref{tab:cfg_analysis:amean_estimates} while in in \cref{fig:cfg_analysis:corr_func_fit}  examples of the correlation function fits are presented.
The final results $\CUSTOMaexpmean$ are means over all $\CUSTOMpdef$ and $L \geq \CUSTOMLmin$ which were chosen for each $\CUSTOMaexp$ according to the quality of the fits.
Please note that we will still refer to different ensembles by the imposed $\CUSTOMaexp$ for clarity.

\begin{table}[t]
	\centering
	\small
	\caption[]{Measured correlation exponents $\CUSTOMaexpmean$ averaged over all concentrations of defects $\CUSTOMpdef$.
		The averages were taken only over $\CUSTOMaexpmean(\CUSTOMpdef,L)$ with $L \geq \CUSTOMLmin(\CUSTOMaexp)$.
		$\nu = 2 / \CUSTOMaexpmean$ are the critical exponent estimates according to the extended Harris criterion.}
	\label{tab:cfg_analysis:amean_estimates}
	$\begin{array}{llll}
\toprule
 \CUSTOMaexp & \CUSTOMaexpmean & 2 / \CUSTOMaexpmean & \CUSTOMLmin \\
\midrule
 3.5   & 3.30(18)  & -             & 112   \\
 3.0   & 2.910(96) & 0.687(23)     & 96    \\
 2.5   & 2.451(26) & 0.8159(86)    & 80    \\
 2.0   & 1.979(18) & 1.0104(89)    & 64    \\
 1.5   & 1.500(30) & 1.333(26)     & 56    \\
\bottomrule
\end{array}$ \end{table}

\begin{figure}[t]
	\centering
	\includegraphics[scale=1]{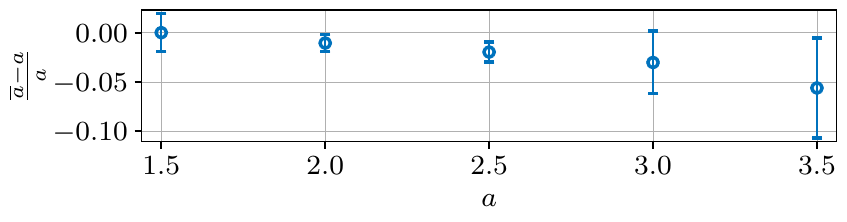}
	\caption[]{Relative deviation of the measured correlation exponents $\CUSTOMaexpmean$ to the imposed values $\CUSTOMaexp$.
		The shown errors of $\CUSTOMaexpmean$ are scaled to $\CUSTOMerrof{\CUSTOMaexpmean} / \CUSTOMaexp$.
		A systematic small underestimation of $\CUSTOMaexp$ can be seen for each $\CUSTOMaexp$.
		It increases with increasing $\CUSTOMaexp$.}
	\label{fig:cfg_analysis:a_mean_deviations}
\end{figure}

As naturally follows from the described determination of $\CUSTOMdistmin$ and $\CUSTOMdistmax$, smaller $\CUSTOMaexpmean(\CUSTOMpdef,L)$ have more degrees of freedom and therefore the estimated values $\CUSTOMaexpmean(\CUSTOMpdef,L)$ coincide better with the proposed $\CUSTOMaexp$.
For weak correlations with $\CUSTOMaexp \geq 3.0$ we exhibit poorer agreement and larger errors for lattice sizes $L \lessapprox 128$.
Also a systematic underestimation of $\CUSTOMaexp$ can be seen in the results.
It becomes more pronounced with larger $\CUSTOMaexp$ and smaller $L$.
We have plotted the relative deviation of the estimates $\CUSTOMaexpmean$ to the expected values $\CUSTOMaexp$ in \cref{fig:cfg_analysis:a_mean_deviations}.
One can see a constant increase in the deviations for increasing $\CUSTOMaexp$.
For our largest $\CUSTOMaexp$ value the deviation reaches $\approx\CUSTOMpercent{5}$.
Nevertheless, we can state that we achieve the desired $\CUSTOMaexp$ values within a precision of $\approx\CUSTOMpercent{5}$.
A test involving more realizations considerably improved the results for the weak correlation cases but we wanted to stay with the number of disorder realizations for which the Monte Carlo simulations were performed later.

\section{Finite-Size Scaling Analysis}
\label{sec:critical_expoent_nu}

We will now discuss the extraction of the critical exponent of the correlation length $\nu$ and the confluent correction exponent $\omega$.
For the finite-size scaling analysis we chose the derivative with respect to the inverse temperature $\beta = 1\CUSTOMikT$ of the logarithm of the magnetization $\partial_{\beta}(\ln \CUSTOMavtot{\CUSTOMoabsm})$.
It can be expressed in terms of expectation values as
\begin{align}
	\partial_{\beta}(\ln \CUSTOMavtot{\CUSTOMoabsm}) & = \frac{\CUSTOMopardiff{\beta} \CUSTOMavtot{\CUSTOMoabsm}}{\CUSTOMavtot{\CUSTOMoabsm}} \notag                                       \\
	                                     & = \CUSTOMvol \frac{\CUSTOMavc{\CUSTOMavt{\CUSTOMabs{\CUSTOMom} \CUSTOMoe}} - \CUSTOMavc{\CUSTOMavt{\CUSTOMabs{\CUSTOMom}}  \CUSTOMavt{\CUSTOMoe}}}{\CUSTOMavc{\CUSTOMavt{\CUSTOMabs{\CUSTOMom}}}} \;,
	\label{eq:nu_exponent:dlnm_def}
\end{align}
where $\CUSTOMavt{\cdot}$ denotes the thermal average and $\CUSTOMavc{\cdot}$ the disorder average and $\CUSTOMoe = \CUSTOMoE / \CUSTOMvol$, $\CUSTOMom = \CUSTOMoM / \CUSTOMvol$ are the normalized energy and magnetization, respectively.
Note, that we use the common convention of taking the absolute value of $\CUSTOMom$ to avoid the trivial averaging to zero in the low-temperature phase for finite lattice sizes.
For the sake of clarity,  we will omit the average brackets for the rest of this work and simply write $\CUSTOModlnm = \partial_{\beta}(\ln \CUSTOMavtot{\CUSTOMoabsm})$.
The derivative of the logarithm of the magnetization $\CUSTOModlnm$ is known to diverge at the critical temperature in the thermodynamic limit $L \CUSTOMlimi$. For finite system sizes it hence develops a minimum.
The finite-size scaling behavior up to the first-order correction reads
\begin{align}
	\CUSTOModlnm_{\min}(L) = \CUSTOMfitfac L^{1/\nu} \CUSTOMrbrl{1 + \CUSTOMfitfactwo L^{-\omega}}\;,
	\label{eq:nu_exponent:dlnm_fss_first_order_corr}
\end{align}
where $\CUSTOModlnm_{\min}(L)$ is the (finite) minimum value of $\CUSTOModlnm(\beta)$ for a given lattice size $L$.
Fitting with this ansatz is difficult as it is a non-linear four-parameter fit.
Therefore, we first determined the correction exponent $\omega$ separately and used it as a fixed parameter in the final estimation.

The whole finite-size scaling analysis can be split into three main steps.
In the first step we derive the peaks of $\CUSTOModlnm$.
The second step is the extraction of the correction exponent $\omega$ which is needed for the fits in the last step.
The last step is the fitting of $\CUSTOModlnm_{\min}(L)$ with fixed $\omega$ and the extraction of $\nu$.

\subsection{Peaks of Observables}
\label{sec:nu_exponent:peak_estimation_dlnm}

We start the analysis with the extraction of the peaks of the derivative of the logarithm of the magnetization $\CUSTOModlnm_{\min}(L)$.
Out of all simulated temperatures for each parameter tuple $(\CUSTOMaexp, \CUSTOMpdef, L)$ we chose three temperatures $\CUSTOMbetasim^i$ with $i = 1,2,3$  where the derivative of the logarithm of the magnetization calculated at these temperatures $\CUSTOModlnm(\CUSTOMbetasim^i)$ was minimal.
For these three $\CUSTOMbetasim^i$ we performed a single histogram reweighting of $\CUSTOModlnm$ to find the minimum values $\CUSTOModlnm^i_{\min}$ and the corresponding temperatures $\beta_{\min}^i$.
The final $\CUSTOModlnm_{\min}$ was chosen to be the minimum of all three $\CUSTOModlnm^i_{\min}$ values.
A more detailed explanation of the reweighting and error estimation process through resampling is presented in \cref{sec:appendix:obs_peak_estimation}.

\begin{figure}[t]
	\centering
	\includegraphics[scale=1]{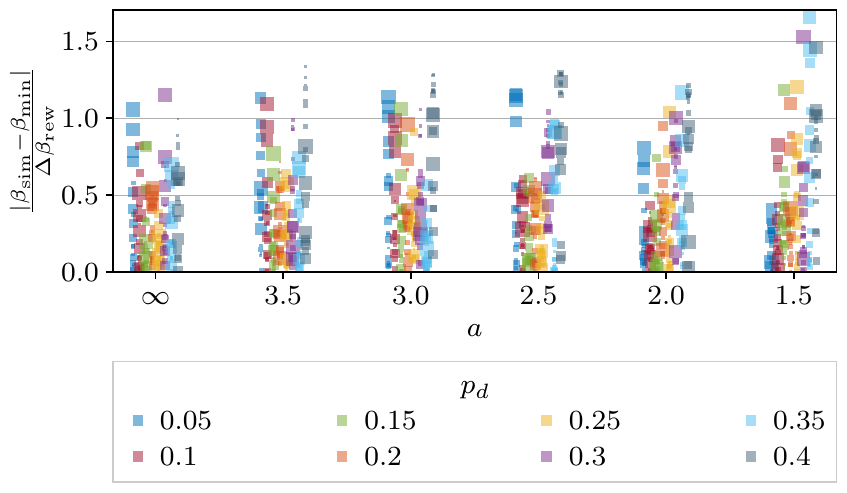}
	\caption[]{The ratios of the differences between simulation temperatures $\CUSTOMbetasim$ and the found temperatures of the minimum values $\beta_{\min}$ with respect to the reweighting range $\CUSTOMobetarew$.
		Smaller points represent smaller lattice sizes $L$.
		The majority of the ratios is $\lessapprox 1$ which is a verification of the reweighting technique validity.
		The maximum values of $\approx 1.5$ are still acceptable.}
	\label{fig:nu_exponent:betarew_distances}
\end{figure}

An important issue was to ensure that the histogram reweighting results lay within the reweighting range.
This is an inevitable restriction coming from the limited statistics.
We used the reweighting range approximation as defined in \autocite{janke2008}
\begin{align}
	\CUSTOMobetarew = \CUSTOMoneover{\sqrt{\CUSTOMavc{\CUSTOMavt{\CUSTOMoE^2}} - \CUSTOMavc{\CUSTOMavt{\CUSTOMoE}^2}}} \;.
	\label{eq:nu_exponent:rew_range_def}
\end{align}
We looked at the ratios of the differences between the simulation temperatures $\CUSTOMbetasim$ and the found temperatures of the minimum values $\beta_{\min}$ with respect to the reweighting range $\CUSTOMobetarew$
\begin{align}
	\frac{\CUSTOMabs{\CUSTOMbetasim - \beta_{\min}}}{\CUSTOMobetarew} \;.
	\label{eq:nu_exponent:betasim_betamin_betarew_ratio}
\end{align}
As can be seen in \cref{fig:nu_exponent:betarew_distances} all obtained $\beta_{\min}$ were close enough to the corresponding $\CUSTOMbetasim$ to assume that the use of the reweighting technique is valid.

\subsection{Confluent Correction Exponent \texorpdfstring{$\omega$}{ω}}
\label{sec:nu_exponent:corr_exponent_omega}

The quotient method which we used for the determination of the confluent correction exponent $\omega$ was successfully used in other works, \CUSTOMeg{} \autocite{ballesteros1998,ballesteros1998a,fytas2016a}.
Starting from an observable $\CUSTOMO$ which has a peak at the critical temperature we build quotients of $\CUSTOMO$ at different lattice sizes $L$ and $sL$
\begin{align}
	\CUSTOMoratio_{\CUSTOMO}(s L) = \frac{\CUSTOMO(s L, \CUSTOMbetac(sL))}{\CUSTOMO(L, \CUSTOMbetac(L))} \;,
	\label{eq:nu_exponent:ratio_def}
\end{align}
where the observables are taken at the critical temperatures for the given lattice sizes $L$ and $sL$, respectively, and $s$ is an arbitrary positive (integer) factor.
For a dimensional observable, \CUSTOMeg{} $\CUSTOModlnm$, the finite-size scaling of $\CUSTOMoratio_{\CUSTOMO}$ in leading order reads \autocite{fytas2016a}
\begin{align}
	\CUSTOMoratio_{\CUSTOMO}(L)  = s^{x_{\CUSTOMO} / \nu} + \CUSTOMfitfac L^{-\omega} \;,
	\label{eq:nu_exponent:ratio_fss_dimensional}
\end{align}
where $x_{\CUSTOMO}$ is the critical exponent of $\CUSTOMO$.

\begin{figure}[t]
	\centering
	\includegraphics[scale=1]{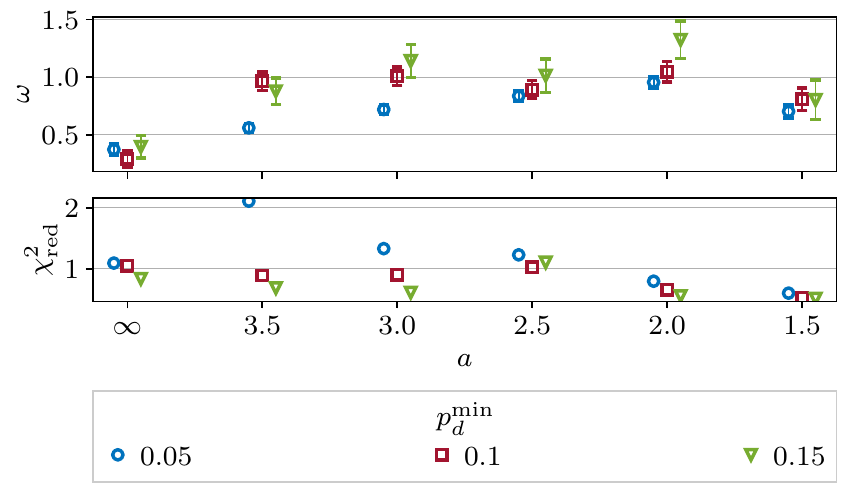}
	\caption[]{Confluent correction exponents $\omega$ from the fits of the quotients to the ansatz $\CUSTOMoratio_{\CUSTOModlnm}(L, \CUSTOMpdef)  = \CUSTOMfitconst + \CUSTOMfitfac_{\CUSTOMpdef} L^{-\omega}$ for all $\CUSTOMaexp$ and various $\CUSTOMpdef^{\min}$.
		The largest included concentration of defects is $\CUSTOMpdef^{\max} = 0.4$.}
	\label{fig:nu_exponent:all_omega}
\end{figure}

\begin{figure}[t]
	\centering
	\subfloat[$\CUSTOMaexp = \infty$.]{\includegraphics[scale=1]{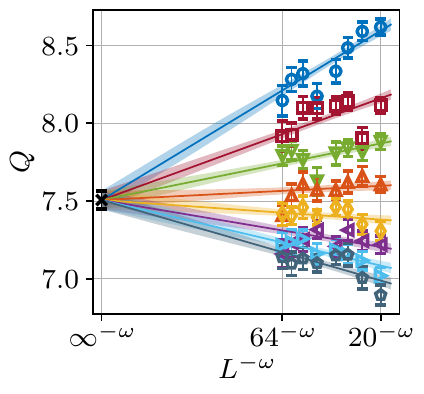}}
	\subfloat[$\CUSTOMaexp = 3.5$.]{\includegraphics[scale=1]{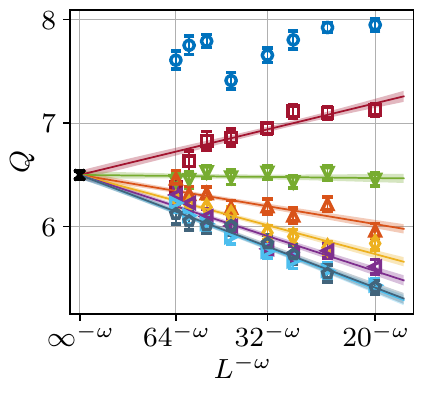}}

	\subfloat[$\CUSTOMaexp = 3.0$.]{\includegraphics[scale=1]{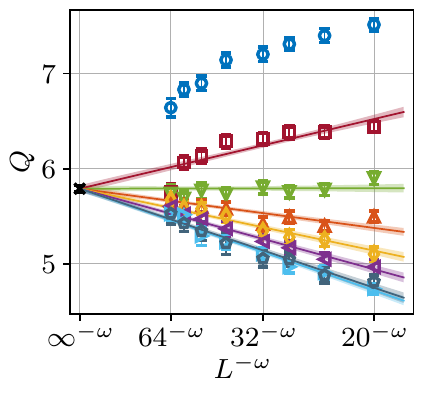}}
	\subfloat[$\CUSTOMaexp = 2.5$.]{\includegraphics[scale=1]{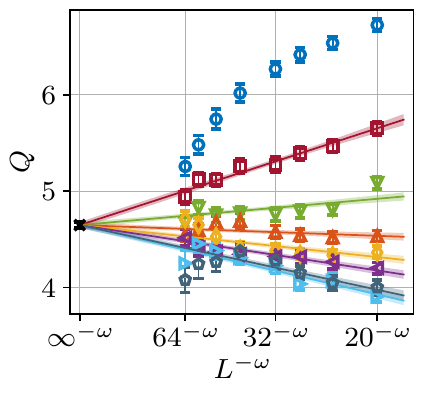}}

	\subfloat[$\CUSTOMaexp = 2.0$.]{\includegraphics[scale=1]{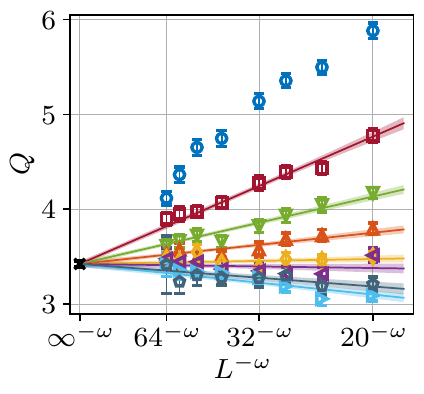}}
	\subfloat[$\CUSTOMaexp = 1.5$.]{\includegraphics[scale=1]{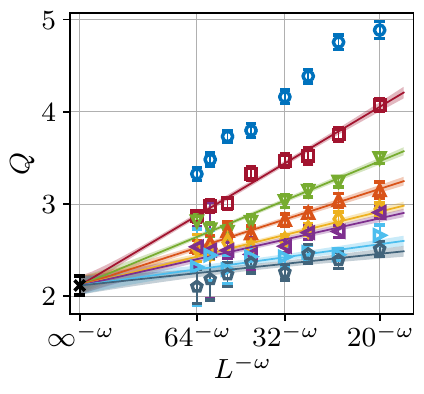}}

	\subfloat{\includegraphics[scale=1]{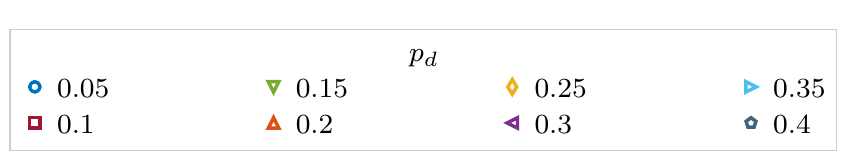}}

	\caption[]{Fits of the quotients of $\CUSTOMoratio_{\CUSTOModlnm}$ at different lattice sizes to the ansatz $\CUSTOMoratio_{\CUSTOModlnm}(L, \CUSTOMpdef)  = \CUSTOMfitconst + \CUSTOMfitfac_{\CUSTOMpdef} L^{-\omega}$ for all $\CUSTOMaexp$.
		The included concentrations for all correlated cases are $0.1 \leq \CUSTOMpdef \leq 0.4$ and $0.05 \leq \CUSTOMpdef \leq 0.4$ for the uncorrelated case.}
	\label{fig:nu_exponent:omega_fits_all_a}
\end{figure}

We calculated the quotients defined through \cref{eq:nu_exponent:ratio_def} for $\CUSTOMO = \CUSTOModlnm$ with $x_\CUSTOMO = 1$ and for $s = 4$.
This allowed us to have 8 independent $\CUSTOMoratio$ values without using the same lattice size twice.
We used the peak values $\CUSTOModlnm_{\min}(L)$ and performed a global fit to $\CUSTOMoratio_{\CUSTOModlnm}(L, \CUSTOMpdef)$ according to \cref{eq:nu_exponent:ratio_fss_dimensional} but using all $\CUSTOMpdef$ simultaneously
\begin{align}
	\CUSTOMoratio_{\CUSTOModlnm}(L, \CUSTOMpdef)  = \CUSTOMfitconst + \CUSTOMfitfac_{\CUSTOMpdef} L^{-\omega} \;,
	\label{eq:nu_exponent:ratio_fss_all_pd_ansatz}
\end{align}
where we explicitly denote the dependence of the amplitudes $\CUSTOMfitampl_{\CUSTOMpdef}$ on the concentrations of defects with the index $\CUSTOMpdef$ and relate the constant $\CUSTOMfitconst$ to \cref{eq:nu_exponent:ratio_fss_dimensional} with
\begin{align}
	\CUSTOMfitconst = s^{x_\CUSTOMO / \nu} \;.
	\label{eq:nu_exponent:ratio_const_relation}
\end{align}
In \cref{fig:nu_exponent:all_omega} we present the $\omega$ results and the qualities of the fits $\CUSTOMchisr$ for $\CUSTOMpdef^{\max} = 0.4$, $\CUSTOMLmin = 20$ and various $\CUSTOMpdef^{\min}$ while in \cref{fig:nu_exponent:omega_fits_all_a} the fits are shown.
$\CUSTOMpdef^{\min}$ and $\CUSTOMpdef^{\max}$ denote the minimum and maximum concentrations of defects included in the fits, respectively.
We have checked the possibility of getting $\omega$ from individual $\CUSTOMpdef$ values but the ratio data suffer from large error bars and the results were not representative.
This fact emphasizes the advantage of using a global fit by simulating at many different concentrations $\CUSTOMpdef$.
Looking into \cref{fig:nu_exponent:all_omega} we see that all fits with $\CUSTOMpdef^{\min} \geq 0.1$ are in a good region of $\CUSTOMchisr \approx 1$ and therefore we took this value as the final values for all correlated cases $\CUSTOMaexp \neq \infty$.
For the uncorrelated case we chose $\CUSTOMpdef^{\min} = 0.05$.
The final $\omega$ results are summarized in \cref{tab:nu_exponent:all_omega}.

\begin{table}[t]
	\centering
	\caption[]{Final confluent correction exponents $\omega$ and constants $\CUSTOMfitconst$ from the fits of the quotients to the ansatz $\CUSTOMoratio_{\CUSTOModlnm}(L, \CUSTOMpdef)  = \CUSTOMfitconst + \CUSTOMfitfac_{\CUSTOMpdef} L^{-\omega}$ for all $\CUSTOMaexp$.
		The maximum included concentration of defects is $\CUSTOMpdef^{\max} = 0.4$.
		As a cross-check we have listed the critical exponents $\nu$ which follow from the relation in \cref{eq:nu_exponent:ratio_const_relation} with $s = 4$.
		They coincide with the final estimates listed in \cref{tab:nu_exponent:nu_exp_fss_results} within the errors.}
	\label{tab:nu_exponent:all_omega}
	\small
	$\begin{array}{llllll}
\toprule
 \CUSTOMaexp  & \omega    & \CUSTOMfitconst & \nu = \frac{\ln s}{\ln \CUSTOMfitconst} & \CUSTOMpdef^{\min} & \CUSTOMchisr \\
\midrule
 \infty & 0.373(53) & 7.506(59) & 0.688(3)                          & 0.05         & 1.095  \\
 3.5    & 0.965(80) & 6.498(40) & 0.741(3)                          & 0.1          & 0.892  \\
 3.0    & 1.008(79) & 5.790(35) & 0.789(3)                          & 0.1          & 0.901  \\
 2.5    & 0.891(79) & 4.648(32) & 0.902(4)                          & 0.1          & 1.026  \\
 2.0    & 1.047(90) & 3.425(33) & 1.126(9)                          & 0.1          & 0.656  \\
 1.5    & 0.808(97) & 2.12(11)  & 1.8(2)                            & 0.1          & 0.529  \\
\bottomrule
\end{array}$ \end{table}

From \cref{fig:nu_exponent:omega_fits_all_a} we can clearly see a distinction between the uncorrelated and correlated cases.
The correction exponent for the uncorrelated case $\omega = \num{0.373(53)}$ matches the prediction $\omega = 0.37(6)$ made by \CUSTOMciteauthorref{ballesteros1998}.
The correction exponent $\omega = \num{1.047(90)}$ for the case $\CUSTOMaexp = 2.0$ is in good agreement with the value $\omega = 1.01(13)$ obtained by \CUSTOMciteauthorref{ballesteros1999}.
A value around $\omega \approx 0.95(10)$ is also found for all other $\CUSTOMaexp$ parameters.
As the errors $\CUSTOMerrof{\omega}$ are quite large for all correlated cases, $\CUSTOMaexp \neq \infty$, chances are that the correction exponent $\omega$ does not depend on $\CUSTOMaexp$ and has a value of roughly $\omega \approx 1$.
Visually it can be verified in \cref{fig:nu_exponent:omega_fits_all_a}.

\subsection{Critical Exponent \texorpdfstring{$\nu$}{ν}}
\label{sec:nu_exponent:crit_exp_nu_estimation}

\begin{figure}[t]
	\centering
	\subfloat[$\CUSTOMaexp = \infty$, $\CUSTOMLmin = 20$.]{\includegraphics[scale=1]{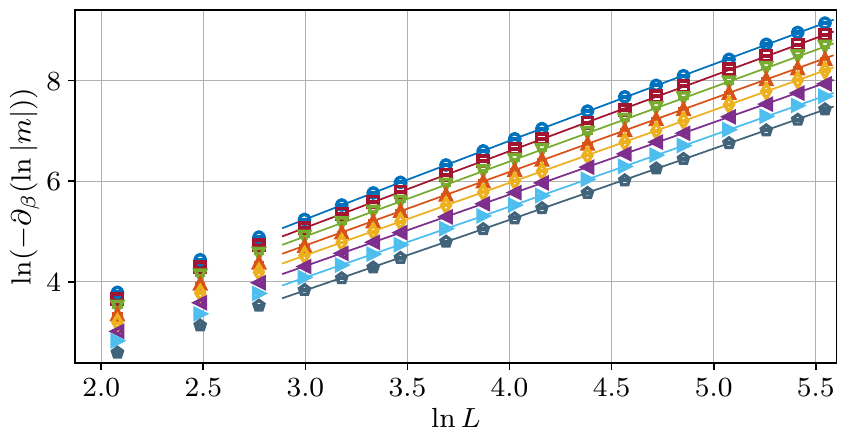}}

	\subfloat[$\CUSTOMaexp = 2.0$, $\CUSTOMLmin = 32$.]{\includegraphics[scale=1]{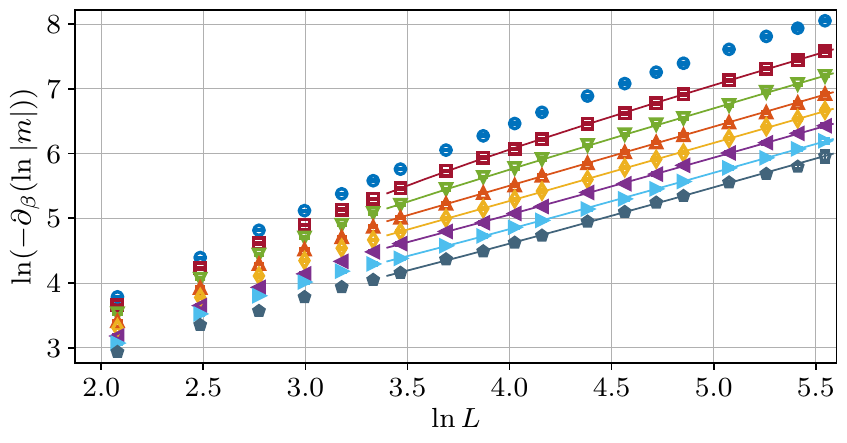}}

	\subfloat{\includegraphics[scale=1]{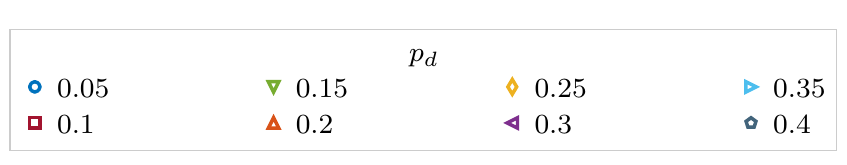}}

	\caption[]{Global fits (solid lines) to the first-order corrected ansatz $\CUSTOModlnm(L, \CUSTOMpdef) = \CUSTOMfitfac_{\CUSTOMpdef} L^{1/\nu} \CUSTOMrbrl{1 + \CUSTOMfitfactwo_{\CUSTOMpdef} L^{-\omega}}$, \cref{eq:nu_exponent:dlnm_fss_first_order_corr_all_pd}, for two different $\CUSTOMaexp$ values.
		For the uncorrected disorder case we used $\CUSTOMpdef^{\min} = 0.05$ and for the correlated cases we used $\CUSTOMpdef^{\min} = 0.1$.}
	\label{fig:nu_exponent:dlnm_corr_fss_dlnm_of_L}
\end{figure}

While the amplitudes $\CUSTOMfitfac$ and $\CUSTOMfitfactwo$ in \cref{eq:nu_exponent:dlnm_fss_first_order_corr} generally depend on $\CUSTOMaexp$ and $\CUSTOMpdef$, $\nu$ and $\omega$ are universal across all $\CUSTOMpdef$ and only show possible dependence on $\CUSTOMaexp$.
This allows us to perform a global fit for each $\CUSTOMaexp$ including all of the $\CUSTOMpdef$ values simultaneously
\begin{align}
	\CUSTOModlnm_{\min}(L, \CUSTOMpdef) = \CUSTOMfitfac_{\CUSTOMpdef} L^{1/\nu} \CUSTOMrbrl{1 + \CUSTOMfitfactwo_{\CUSTOMpdef} L^{-\omega}}\;,
	\label{eq:nu_exponent:dlnm_fss_first_order_corr_all_pd}
\end{align}
where we explicitly denoted the dependence of $\CUSTOMfitfac_{\CUSTOMpdef}$ and $\CUSTOMfitfactwo_{\CUSTOMpdef}$ on $\CUSTOMpdef$.
We performed least squares fits to \cref{eq:nu_exponent:dlnm_fss_first_order_corr_all_pd} with $\CUSTOMfitfac_{\CUSTOMpdef}$, $\CUSTOMfitfactwo_{\CUSTOMpdef}$ and $1/\nu$ as parameters and used fixed correction exponents $\omega(\CUSTOMaexp)$ listed in \cref{tab:nu_exponent:all_omega}.
Examples of the resulting fits are shown in \cref{fig:nu_exponent:dlnm_corr_fss_dlnm_of_L}.

\begin{figure}[t]
	\centering
	\subfloat[$\CUSTOMaexp = \infty$.]{\includegraphics[scale=1]{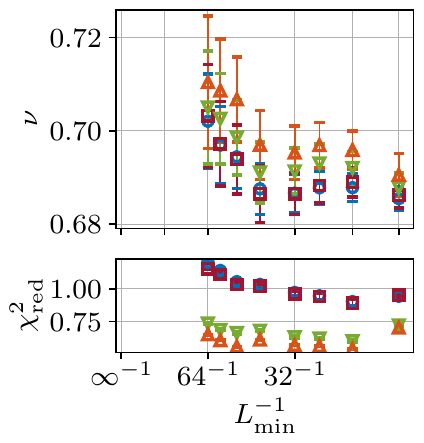}}
	\subfloat[$\CUSTOMaexp = 3.5$.]{\includegraphics[scale=1]{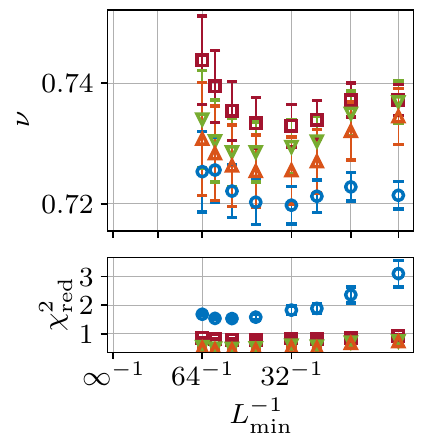}}

	\subfloat[$\CUSTOMaexp = 3.0$.]{\includegraphics[scale=1]{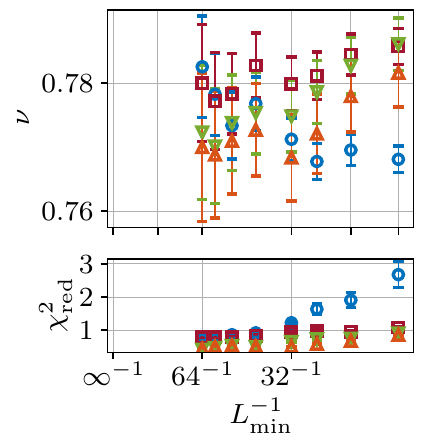}}
	\subfloat[$\CUSTOMaexp = 2.5$.]{\includegraphics[scale=1]{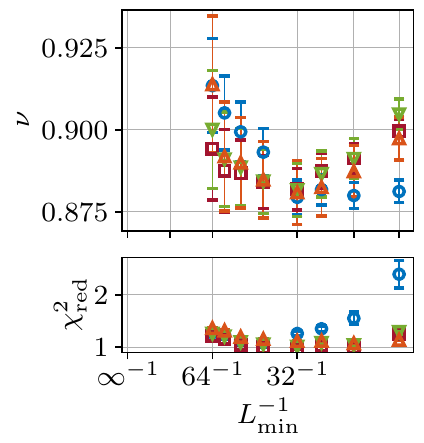}}

	\subfloat[$\CUSTOMaexp = 2.0$.]{\includegraphics[scale=1]{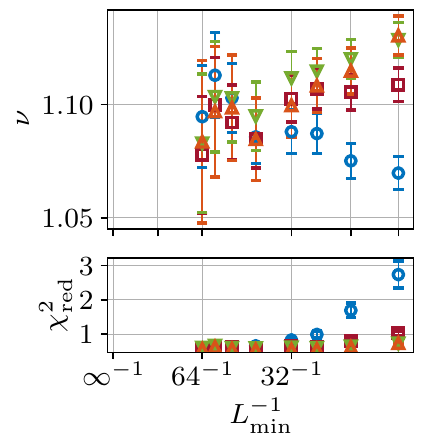}}
	\subfloat[$\CUSTOMaexp = 1.5$.]{\includegraphics[scale=1]{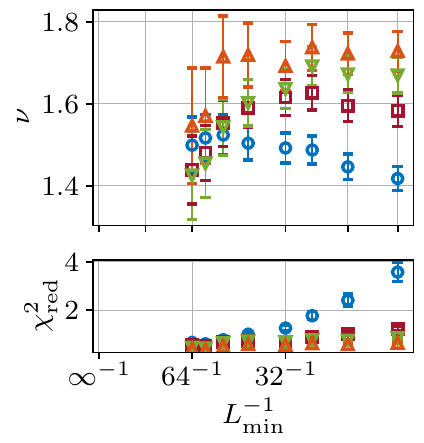}}

	\subfloat{\includegraphics[scale=1]{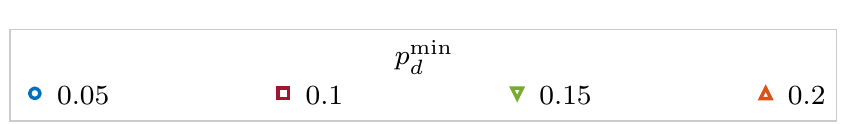}}

	\caption[]{Critical exponents $\nu$ from fits to the first-order corrected ansatz $\CUSTOModlnm(L, \CUSTOMpdef) = \CUSTOMfitfac_{\CUSTOMpdef} L^{1/\nu} \CUSTOMrbrl{1 + \CUSTOMfitfactwo_{\CUSTOMpdef} L^{-\omega}}$, \cref{eq:nu_exponent:dlnm_fss_first_order_corr_all_pd}, with $\CUSTOMpdef^{\max} = 0.4$ and varying $\CUSTOMpdef^{\min}$ and $\CUSTOMLmin$.
		The corresponding $\CUSTOMchisr$ are shown as a second plot for each $\CUSTOMaexp$.
		For $\CUSTOMaexp \leq 2.5$ one sees a dependence on $\CUSTOMLmin$ and $\CUSTOMpdef^{\min}$.
		For larger $\CUSTOMaexp$ the errors usually overlap for each $\CUSTOMpdef^{\min}$ and also the dependence on $\CUSTOMLmin$ is mainly covered by the error sizes which become larger for larger $\CUSTOMLmin$.
	}
	\label{fig:nu_exponent:nu_fit_Lmin_pd_min}
\end{figure}

\begin{figure}[t]
	\centering
	\includegraphics[scale=1]{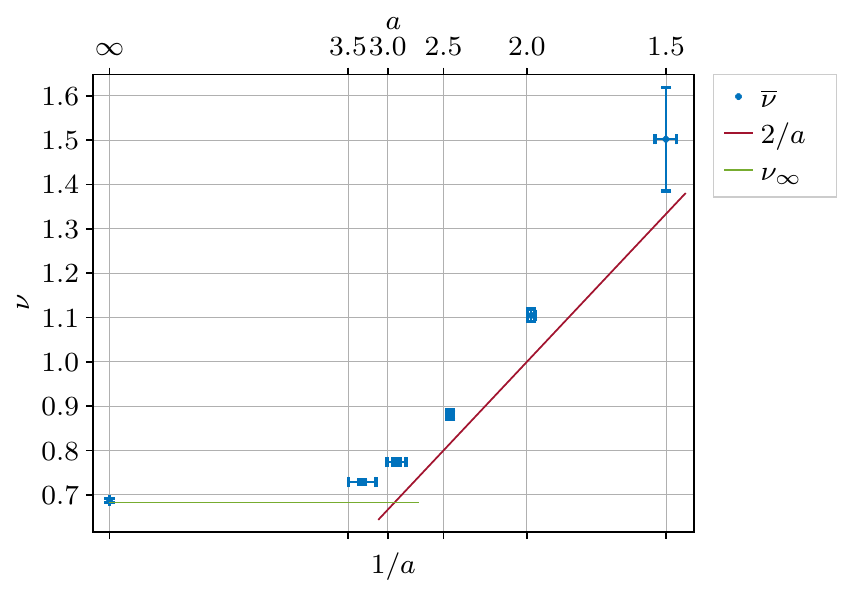}
	\caption[]{Final Results of the critical exponents $\CUSTOMmean{\nu}$ plotted over $1 / \CUSTOMaexp$.
		The chosen concentration limits were $\CUSTOMpdef^{\min} = 0.1$ for the correlated disorder cases and $\CUSTOMpdef^{\min} = 0.05$ for the uncorrelated disorder case and $\CUSTOMpdef^{\max} = 0.4$.
		Horizontal errors are errors of measured $\CUSTOMaexpmean$ listed in \cref{tab:cfg_analysis:amean_estimates} and scaled to $1 / \CUSTOMaexp$.
		The uncorrelated disorder case critical exponent was set to $\nu_\infty = 0.683$ as an average value from other works listed in \cref{tab:introduction:disorder_nu_literature}.
		Expected values $\nu = 2/\CUSTOMaexpmean$ according to the prediction of the extended Harris criterion are shown for comparison for all $\CUSTOMaexp \leq \CUSTOMdim$ where the extended Harris criterion is assumed to be valid.}
	\label{fig:nu_exponent:nu_exp_fss_results}
\end{figure}

We performed the fits for various minimal lattice sizes $20 \leq \CUSTOMLmin \leq 64$.
We also varied the smallest concentration $\CUSTOMpdef^{\min}$ and the largest concentration $\CUSTOMpdef^{\max}$ included into the global fit.
The variation of $\CUSTOMpdef^{\max}$ turned out to be neglectable and we finally chose $\CUSTOMpdef^{\max} = 0.4$.
The dependency of the resulting critical exponent $\nu$ on $\CUSTOMLmin$ and $\CUSTOMpdef^{\min}$ is shown in \cref{fig:nu_exponent:nu_fit_Lmin_pd_min} for all $\CUSTOMaexp$.
The deviation of the fit results for $\CUSTOMpdef^{\min} = 0.05$ from all other cases $\CUSTOMpdef^{\min} > 0.05$ was significant for all correlated cases.
Additionally the goodness of the fits $\CUSTOMchisr$ was poor in these cases.
When the $\CUSTOMpdef = 0.05$ data sets were excluded, the fits showed good behavior.
We chose $\CUSTOMpdef^{\min} = 0.1$ for final estimates for the correlated disorder cases and left $\CUSTOMpdef^{\min} = 0.05$ for the uncorrelated case.
However, in order to further take into account the deviations of the results for different $\CUSTOMpdef^{\min}$ we took the smallest $\CUSTOMLmin$ parameter where the errors of the fits for different $\CUSTOMpdef^{\min}$ mostly overlapped for the first time.
The final $\CUSTOMLmin$ parameters and the corresponding $\CUSTOMchisr$ as well as the final estimated critical exponents $\CUSTOMmean{\nu}$ are listed in \cref{tab:nu_exponent:nu_exp_fss_results}.
Additionally, the $\CUSTOMmean{\nu}$ values are shown in \cref{fig:nu_exponent:nu_exp_fss_results}.

\begin{table}[t]
	\centering
	\caption[]{Final results of the critical exponents $\nu$.
		The chosen concentration limits were $\CUSTOMpdef^{\min} = 0.1$ for the correlated cases and $\CUSTOMpdef^{\min} = 0.05$ for the uncorrelated case and $\CUSTOMpdef^{\max} = 0.4$.
		Expected values $\nu = 2/\CUSTOMaexpmean$ according to the prediction of the extended Harris criterion are shown for comparison for all $\CUSTOMaexp \leq \CUSTOMdim$ where the extended Harris criterion is assumed to be valid.
		For completeness the correction exponents $\omega$ from \cref{tab:nu_exponent:all_omega} are listed once again.}
	\label{tab:nu_exponent:nu_exp_fss_results}
	\small
	$\begin{array}{llllll}
\toprule
 \CUSTOMaexp  & \CUSTOMmean{\nu} & 2 / \CUSTOMaexpmean & \CUSTOMchisr     & \CUSTOMLmin & \omega    \\
\midrule
 \infty & 0.6875(47) & -             & 0.829(17)  & 20    & 0.373(53) \\
 3.5    & 0.7293(56) & -             & 0.650(27)  & 32    & 0.965(80) \\
 3.0    & 0.7744(68) & 0.687(23)     & 0.700(30)  & 32    & 1.008(79) \\
 2.5    & 0.8814(99) & 0.8159(86)    & 1.0315(33) & 32    & 0.891(79) \\
 2.0    & 1.105(15)  & 1.0104(89)    & 0.6055(72) & 32    & 1.047(90) \\
 1.5    & 1.50(12)   & 1.333(26)     & 0.4242(65) & 56    & 0.808(97) \\
\bottomrule
\end{array}$ \end{table}

The obtained value for the uncorrelated case \mbox{$\nu = \num{0.6875(47)}$} is in very good agreement with the results from other groups listed in \cref{tab:introduction:disorder_nu_literature}.
Please note that in most works the $\nu$ exponent was concentration dependent in contrast to the present work.
Therefore the comparison must be done with care.
Altogether we can conclude that our extraction method and in particular the global fit ansatz work well for the uncorrelated case which can be seen as a good verification.

For the correlated disorder cases we first compare our results to the prediction of the extended Harris criterion.
All obtained values lie about \CUSTOMpercent{10} above the prediction of $\nu = 2 / \CUSTOMaexp$.
Nevertheless we see the right tendency of the $\nu$ values in being proportional to $1 / \CUSTOMaexp$ and in approaching the uncorrelated case somewhere around $\CUSTOMaexp \approx 3.0$.
The crossover region around $\CUSTOMaexp \approx 3.0$ shows the largest deviations from the extended Harris criterion estimate as well as from the uncorrelated case.
This behavior is expected for finite systems.
The estimate for $\CUSTOMaexp = 1.5$ has a huge error and therefore is not very representative.
Probably more realizations are needed to get a better result for such strongly correlated case.

Considering the $\nu$ values for the correlated case with $\CUSTOMaexp = 2.0$ we see a discrepancy between our results and results from other groups listed in \cref{tab:introduction:disorder_nu_literature} (see also the summary plot in \cref{fig:conclusions:nu_exp_results_literature}).
There are several possible reasons for such deviations.
Comparing to the work of \CUSTOMciteauthorsref{ballesteros1999} and \CUSTOMciteauthorsref{ivaneyko2008}, we used much more finer lattice size stepping; 18 lattice sizes in the range of $8 \leq L \leq 256$ versus 5 sizes in the range $8 \leq L \leq 128$.
The number of realizations in our case was smaller by a factor of 10 but we measured 10 times longer time series on each realization.
Further, we used the derivative of the logarithm of the magnetization $\CUSTOModlnm$ as our primary observable whereas in the other works the derivatives of Binder cumulants $\CUSTOModutwo$ and $\CUSTOModufour$ were used.
Additionally, the concrete methods of generating the correlated disorder and extracting the critical exponent $\nu$ were very different.
Finally, but probably most importantly, the method used in this work included all $\CUSTOMpdef$ values in the critical exponent $\nu$ estimation.
Comparing our $\nu$ exponents to the results of \CUSTOMciteauthorref{prudnikov2000} and \autocite{prudnikov2005} we do not see any agreement.
The reason for this remains unclear to us.

\subsection{Critical Temperature}
\label{sec:temperature}

\begin{figure}[t]
	\centering
	\includegraphics[scale=1]{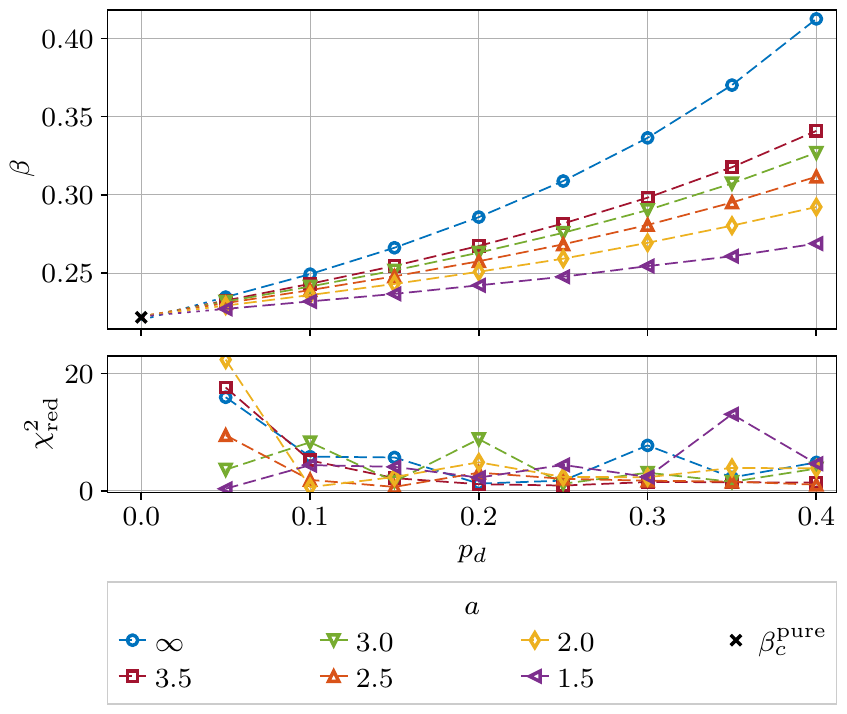}
	\caption[]{Critical temperatures $\CUSTOMbetac$ obtained from fits to the ansatz $\beta_{\min}(L) = \CUSTOMbetac + \CUSTOMfitampl L^{-1/\nu}$, \cref{eq:temperature:betac_fss_relation}, for all simulated correlation exponents $\CUSTOMaexp$ and concentrations of defects $\CUSTOMpdef$.
		The dashed lines are shown to guide the eyes.
		For the extension to $\CUSTOMpdef = 0$ we extrapolated the lines connecting the points at $\CUSTOMpdef = 0.05$ and $\CUSTOMpdef = 0.1$.
		This was done for a visual control of how the curves approximately approach the pure case limit.
		The critical temperature for the pure case was set to $\beta_{\CUSTOMcrit}^{\CUSTOMpure} = \num{0.221654628(2)}$ from Ref.~\autocite{ferrenberg2018}.
	}
	\label{fig:temperature:crit_temp_and_chi_dlnm}
\end{figure}

\begin{figure}[t]
	\centering
	\subfloat[$\CUSTOMaexp = 2.0$, $\CUSTOMpdef = 0.25$.]{\includegraphics[scale=1]{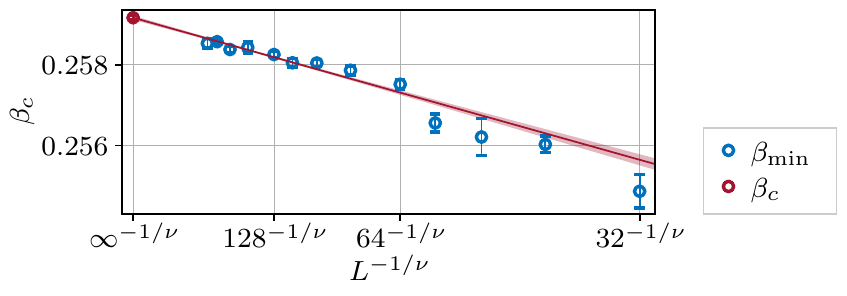}}

	\subfloat[$\CUSTOMaexp = \infty$, $\CUSTOMpdef = 0.25$.]{\includegraphics[scale=1]{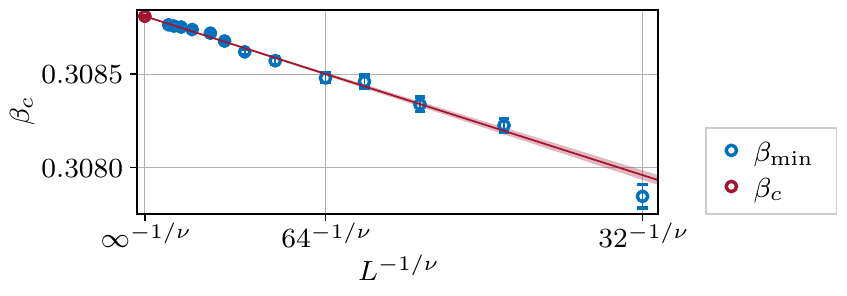}}

	\caption[]{Fits of $\beta_{\min}(L)$ corresponding to the minima of $\CUSTOModlnm$ to the ansatz $\beta_{\min}(L) = \CUSTOMbetac + \CUSTOMfitampl L^{-1/\nu}$, \cref{eq:temperature:betac_fss_relation}, for different correlation exponents $\CUSTOMaexp$ and $\CUSTOMpdef = 0.25$.
		The correlation length critical exponents $\CUSTOMmean{\nu}$ are taken from \cref{tab:nu_exponent:nu_exp_fss_results}.}
	\label{fig:temperature:crit_temp_fss_fits_dlnm}
\end{figure}

Once we have derived the peaks of $\CUSTOModlnm$ in \cref{sec:nu_exponent:peak_estimation_dlnm}, we also had the corresponding temperatures $\beta_{\min}$.
This allowed us to study the critical temperatures for for all correlation exponents $\CUSTOMaexp$ and concentrations of defects $\CUSTOMpdef$.
Note, that unlike for the critical exponent $\nu$ we need to attend each $\CUSTOMpdef$ separately and cannot perform a global fit as the critical temperature depends on it.
To obtain the critical temperatures $\CUSTOMbetac$ for all $\CUSTOMaexp$ and $\CUSTOMpdef$ values we used the finite-size scaling relation in the leading order
\begin{align}
	\beta_{\min}(L) = \CUSTOMbetac + \CUSTOMfitampl L^{-1/\nu} \;,
	\label{eq:temperature:betac_fss_relation}
\end{align}
where $\beta_{\min}$ are the temperatures corresponding to the minimal values of the derivative of the logarithm of magnetization $\CUSTOModlnm_{\min}$ at different $L$ and $\CUSTOMbetac$ is the desired critical temperature at $L \CUSTOMlimi$.
We performed the fits to the ansatz given in \cref{eq:temperature:betac_fss_relation} by using the extracted exponents  $\CUSTOMmean{\nu}$ for the corresponding $\CUSTOMaexp$ values listed in \cref{tab:nu_exponent:nu_exp_fss_results}.
The quality of the fits was moderate and varied for different $\CUSTOMpdef$ and $\CUSTOMaexp$ significantly.
Therefore we set $\CUSTOMLmin = 32$ for all parameter tuples.
Finally, to incorporate the uncertainties in the $\CUSTOMmean{\nu}$ estimates we performed the fits in a bootstrapped way by randomly choosing a $\nu_i = \CUSTOMtrm{Normal}(\CUSTOMmean{\nu}, \CUSTOMerrof{\CUSTOMmean{\nu}})$ according to a normal distribution and performing \num{10000} fits.
All final quantities were averages over these bootstrapped fits.
The resulting temperatures and the qualities of the fits are presented in \cref{fig:temperature:crit_temp_and_chi_dlnm,tab:temperature:crit_temp_dlnm_results}.
Examples of the fits for different $\CUSTOMaexp$ and $\CUSTOMpdef$ can be found in \cref{fig:temperature:crit_temp_fss_fits_dlnm}.

\begin{table}[t]
	\small
	\centering
	\caption[]{Critical temperatures $\CUSTOMbetac$ obtained from fits to the ansatz $\beta_{\min}(L) = \CUSTOMbetac + \CUSTOMfitampl L^{-1/\nu}$, \cref{eq:temperature:betac_fss_relation}, for all simulated correlation exponents $\CUSTOMaexp$ and concentrations of defects $\CUSTOMpdef$.
		The corresponding $\CUSTOMchisr$ values are shown in \cref{fig:temperature:crit_temp_and_chi_dlnm}.}
	\label{tab:temperature:crit_temp_dlnm_results}
	\sisetup{
	table-text-alignment=center,
	table-format = 1.8,
}
$\begin{array}{lSSS}
\toprule
 \CUSTOMpdef & {\CUSTOMaexp = \infty} & {\CUSTOMaexp = 3.5} & {\CUSTOMaexp = 3.0} \\
\midrule
 0.05  & 0.234598(2)      & 0.232412(2)   & 0.231737(2)   \\
 0.1   & 0.249289(2)      & 0.243087(3)   & 0.241352(4)   \\
 0.15  & 0.266155(2)      & 0.254596(4)   & 0.251635(6)   \\
 0.2   & 0.285755(3)      & 0.267326(6)   & 0.263032(9)   \\
 0.25  & 0.308812(4)      & 0.281649(9)   & 0.27571(2)    \\
 0.3   & 0.336423(5)      & 0.29818(2)    & 0.29025(2)    \\
 0.35  & 0.370154(7)      & 0.31764(2)    & 0.30729(4)    \\
 0.4   & 0.412487(9)      & 0.34087(3)    & 0.32692(3)    \\
\midrule
 \CUSTOMpdef & {\CUSTOMaexp = 2.5}    & {\CUSTOMaexp = 2.0} & {\CUSTOMaexp = 1.5} \\
\midrule
 0.05  & 0.230755(4)      & 0.229190(9)   & 0.22707(3)    \\
 0.1   & 0.239077(9)      & 0.23592(3)    & 0.23185(4)    \\
 0.15  & 0.24790(2)       & 0.24303(3)    & 0.23665(7)    \\
 0.2   & 0.25753(2)       & 0.25080(4)    & 0.24215(6)    \\
 0.25  & 0.26835(3)       & 0.25917(6)    & 0.24762(8)    \\
 0.3   & 0.28074(4)       & 0.26931(9)    & 0.2544(3)     \\
 0.35  & 0.29508(6)       & 0.28028(9)    & 0.2607(1)     \\
 0.4   & 0.31158(5)       & 0.29222(6)    & 0.2688(2)     \\
\bottomrule
\end{array}$ \end{table}

The qualitative behavior of the temperature curves is in strong agreement with the expectations.
When the concentration of defects vanishes, $\CUSTOMpdef \CUSTOMlimz$, the inverse temperature goes to the pure Ising model case with \mbox{$\CUSTOMbetac = \num{0.221654628(2)}$} \autocite{ferrenberg2018}.
On the other hand, when the concentration approaches the percolation threshold concentration, $\CUSTOMpdef \CUSTOMra \CUSTOMpdefth(\CUSTOMaexp)$, Ref.~\autocite{zierenberg2017}, the inverse temperature becomes infinity, $\CUSTOMbetac \CUSTOMlimi$.
In contrast to the minimal values $\CUSTOModlnm_{\min}$ which were obtained with a high accuracy, it was not possible to get such precise temperatures $\beta_{\min}$.
The main difficulty was the large width of the peaks of $\CUSTOModlnm(\beta)$ for stronger correlations.
Additionally, in some cases the reweighting range was not large enough to cover the temperature of the peak sufficiently.
Nevertheless, the estimates provide a consistent picture and can serve as a good starting point for later analyses.

\section{Conclusions}
\label{sec:conclusions}

We applied Monte Carlo simulation techniques to the three-dimensional Ising model on a lattice with long-ranged correlated site disorder.
The correlation of the disorder was assumed to be proportional to a power-law $\propto \CUSTOMdist^{- \CUSTOMaexp}$ with a correlation exponent $\CUSTOMaexp$.
We provided a decent analysis of the disorder correlation in our disorder ensembles verifying the correlation exponent $\CUSTOMaexp$ numerically.

We found the critical exponents of the correlation length $\nu$ and the confluent correction exponents $\omega$ as well as the critical temperatures $\CUSTOMbetac$ of the system for various correlation exponents $1.5 \leq \CUSTOMaexp \leq 3.5$ as well as for the uncorrelated case $\CUSTOMaexp = \infty$.
Contrarily to other works we performed a global fit where we included different disorder concentrations into one simultaneous fit.
Such a study was not possible before because all known works only considered one particular correlation exponent choice $\CUSTOMaexp = 2.0$ and only one or two different concentrations $\CUSTOMpdef$ whereas in this work we used a wide range of $\CUSTOMaexp$ and $\CUSTOMpdef$ values.

\begin{figure}[t]
	\centering
	\includegraphics[scale=1]{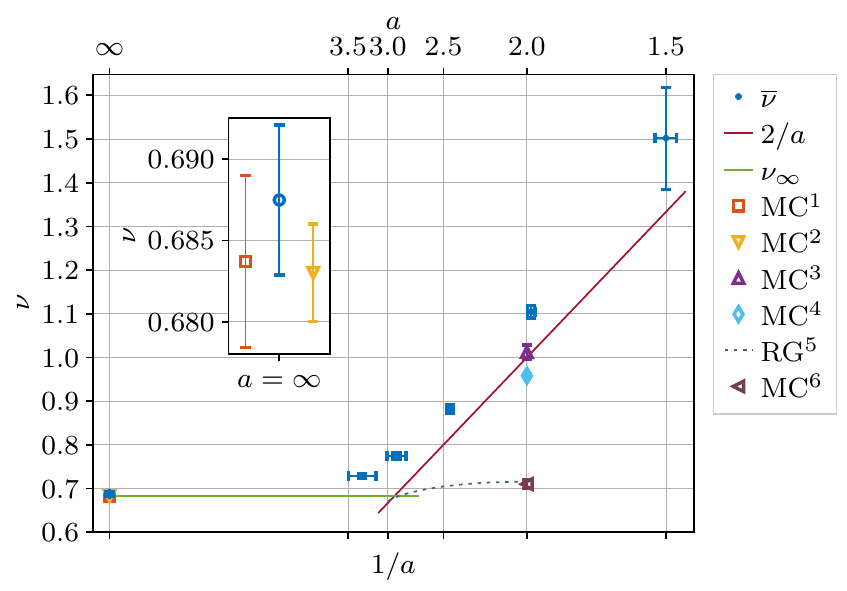}
	\caption[]{Final Results of the critical exponent $\nu$ compared to the known results from the literature and the prediction of the extended Harris criterion $\nu = 2 / \CUSTOMaexp$.
		1: \CUSTOMciteauthorref{ballesteros1998}, 2: \CUSTOMciteauthorref{calabrese2003}, 3: \CUSTOMciteauthorref{ballesteros1999}, 4: \CUSTOMciteauthorref{ivaneyko2008}, 5: \CUSTOMciteauthorref{prudnikov2000}, 6: \CUSTOMciteauthorref{prudnikov2005}.
		The inset shows a close up of the uncorrelated case $\CUSTOMaexp  = \infty$.
		The uncorrelated disorder case critical exponent was set to $\nu_\infty = 0.683$ as an average value from other works listed in \cref{tab:introduction:disorder_nu_literature}.
		The results of this work lie about \CUSTOMpercent{10} above the prediction of the extended Harris criterion $\nu = 2/ \CUSTOMaexp$.
		On the other hand, they also do not coincide with other works.
		The main reason for this discrepancy is probably the global fit ansatz of the present work which combines all $\CUSTOMpdef$ into one single fit.}
	\label{fig:conclusions:nu_exp_results_literature}
\end{figure}

In \cref{fig:conclusions:nu_exp_results_literature} we give a visual comparison of the critical exponents $\nu$ obtained in this work, results known from other works and predictions by the extended Harris criterion.
We obtain a value $\nu = \num{0.6875(47)}$ for the uncorrelated case which matches the results from other groups listed in \cref{tab:introduction:disorder_nu_literature} and plotted in \cref{fig:conclusions:nu_exp_results_literature}.
Also the correction exponent $\omega = \num{0.373(53)}$ coincides with Ref.~\autocite{ballesteros1999}.

The estimated $\nu$ values for the correlated disorder cases show the $1/\CUSTOMaexp$ behavior predicted by the extended Harris criterion qualitatively but are approximately \CUSTOMpercent{10} higher than the prediction $2 / \CUSTOMaexp$.
On the other hand, we strongly disagree with the renormalization group predictions made by \CUSTOMciteauthorref{prudnikov2000} and their Monte Carlo simulation result for $\CUSTOMaexp = 2.0$ in Ref.~\autocite{prudnikov2005}.
The correction exponent $\omega = \num{1.047(90)}$ for the case $\CUSTOMaexp = 2.0$ is in good agreement with \CUSTOMciteauthorref{ballesteros1999}.
For all correlated cases we measure a value which is compatible with $\omega = 0.95(10) \approx 1$.

Our estimation of the critical temperatures $\CUSTOMbetac$ provides a global picture of the system for different $\CUSTOMaexp$ and $\CUSTOMpdef$ parameters and can serve as a good starting point for further analyses and simulations.

In upcoming studies we will consider other critical exponents like $\beta$ and $\gamma$ and hopefully tackle down the problem even more.

\section*{Acknowledgments}
\label{sec:acknowledgments}

The authors would like to thank the Max Planck Society and in particular the Max Planck Institute of Mathematics in the Sciences for financial support of this work and for providing the computational resources at Max Planck Computing and Data Facility.
Further support by the Deutsch-Französische Hochschule (DFH-UFA) through the Doctoral College "$\mathbb{L}^4$" under Grant No.\ CDFA-02-07 is gratefully acknowledged.
Many thanks to Christophe Chatelain, Malte Henkel, Yurij Holovatch and Mikhail Nalimov for intense discussions.

\appendix
\section{Estimation of Peaks of Observables}
\label{sec:appendix}
\label{sec:appendix:obs_peak_estimation}

\newcommand{\CUSTOMOP}{{\mathcal{P}\,}}
\newcommand{\CUSTOMOA}{\mathcal{A}}
\newcommand{\CUSTOMOB}{\mathcal{B}}
\newcommand{\CUSTOMf}{f}
\newcommand{\CUSTOMjki}{j}
\newcommand{\CUSTOMjkc}{k}
\newcommand{\CUSTOMjka}{a}
\newcommand{\CUSTOMJK}{J}
\newcommand{\CUSTOMNjki}{N_{\CUSTOMjki}}
\newcommand{\CUSTOMNjkc}{N_{\CUSTOMjkc}}
\newcommand{\CUSTOMNjka}{N_{\CUSTOMjka}}
\newcommand{\CUSTOMbetamin}{\check{\beta}}
\newcommand{\CUSTOMOPmin}{\check{\mathcal{P}}\,}
\newcommand{\CUSTOMOmin}{\check{\CUSTOMO}}

Suppose we performed simulations on $\CUSTOMncfg$ disorder realizations and did $\CUSTOMnmeas$ measurements at a simulation temperature $\CUSTOMbetasim$ on each of them.
We are equipped with two-dimensional arrays of total energy $\CUSTOMoE_i^c$ and total magnetization $\CUSTOMoM_i^c$ for $i = 1, \dots, \CUSTOMnmeas$ and $c = 1, \dots , \CUSTOMncfg$.
Using these arrays we can calculate observables of the from
\begin{align}
	\CUSTOMO_i^c = (\CUSTOMoE_i^c)^k (\CUSTOMoM_i^c)^l \;,
	\label{eq:appendix:obs_composed_of_E_and_M}
\end{align}
where $k$ and $l$ are arbitrary powers.
We introduce the notation $\CUSTOMO^c$ for an average over $\CUSTOMO_i^c$ for one particular disorder realization $c$
\begin{align}
	\CUSTOMO^c = \CUSTOMavt{\CUSTOMO} = \CUSTOMoneover{\CUSTOMnmeas} \sum_{i = 1}^{\CUSTOMnmeas} \CUSTOMO_i^c \;.
	\label{eq:appendix:thermal_average_of_obs_O}
\end{align}
The average over the disorder realizations is denoted by $\CUSTOMavc{\cdot}$ and the final estimate $\CUSTOMO$ reads
\begin{align}
	\CUSTOMO = \CUSTOMavtot{\CUSTOMO} = \CUSTOMavc{\CUSTOMO^c} = \CUSTOMoneover{\CUSTOMncfg} \sum_{c = 1}^{\CUSTOMncfg} \CUSTOMO^c \;.
	\label{eq:appendix:disorder_average_of_obs_O}
\end{align}

For variables of the type of \cref{eq:appendix:obs_composed_of_E_and_M} a histogram reweighting technique can be used to reweight the observable from the simulated temperature $\CUSTOMbetasim$ to a different temperature $\beta$.
We used the form given in \autocite{janke2008}
\begin{align}
	\CUSTOMrew(\CUSTOMO)^{c}(\beta) = \frac{\sum_{i = 1}^\CUSTOMnmeas \CUSTOMO_i^c \CUSTOMeto{-(\beta - \CUSTOMbetasim) \CUSTOMoE_i^c}}{\sum_{i = 1}^\CUSTOMnmeas \CUSTOMeto{-(\beta - \CUSTOMbetasim)\CUSTOMoE_i^c}} \;,
	\label{eq:appendix:hist_rew_each_disorder_obs_O}
\end{align}
where the reweighting is performed separately for each disorder realization $c$ and the final estimate at the temperature $\beta$ is the disorder average
\begin{align}
	\CUSTOMrew(\CUSTOMO)(\beta) = \CUSTOMavc{\CUSTOMrew(\CUSTOMO)^c} \;.
	\label{eq:appendix:hist_rew_disorder_average_obs_O}
\end{align}

Not every observable of interest, in particular the derivative of the logarithm of the magnetization $\CUSTOModlnm$ has the form of \cref{eq:appendix:obs_composed_of_E_and_M}.
Let $\CUSTOMOP$ denote a composite observable of the following form
\begin{align}
	\CUSTOMOP  = \CUSTOMf(\CUSTOMO^{(1)}, \CUSTOMO^{(2)}, \dots) \;,
	\label{eq:appendix:def_composed_obs_P}
\end{align}
where each of $\CUSTOMO^{(k)}$ fulfills the form of \cref{eq:appendix:obs_composed_of_E_and_M}.
For this kind of composed observables we define the reweighting procedure by reweighting each component separately
\begin{align}
	\CUSTOMrew(\CUSTOMOP)(\beta) = \CUSTOMf(\CUSTOMrew(\CUSTOMO^{(1)})(\beta), \CUSTOMrew(\CUSTOMO^{(2)})(\beta), \dots) \;.
	\label{eq:appendix:hist_rew_composed_obs_P}
\end{align}
Let us summarize what we have achieved so far.
Starting with the arrays of raw observables $\CUSTOMoE$ and $\CUSTOMoM$ we are able to use the histogram reweighting technique to obtain practically any observable calculable from $\CUSTOMoE$ and $\CUSTOMoM$ as a function of $\beta$.

Let us now assume that the finite-size scaling analysis of $\CUSTOMOP(\beta)$ predicts a minimum $\CUSTOMOPmin$ at a certain temperature $\CUSTOMbetamin$.
Without loss of generality we assume a minimum of $\CUSTOMOP(\beta)$, otherwise we transform $\CUSTOMOP \CUSTOMra - \CUSTOMOP$.
In the thermodynamic limit $L \CUSTOMlimi$ we expect $\CUSTOMbetamin \CUSTOMra \CUSTOMbetac$.
We can apply an optimization routine by plugging in $\CUSTOMrew(\CUSTOMOP)(\beta)$ as the target function and obtain $\CUSTOMOPmin$ and $\CUSTOMbetamin$,
\begin{align}
	\CUSTOMOPmin = \min_{\beta} \CUSTOMrbrl{\CUSTOMrew(\CUSTOMOP)(\beta)} \;.
	\label{eq:appendix:hist_rew_dlnm_minimization}
\end{align}

However, we will not be able to estimate the errors $\CUSTOMerrof{\CUSTOMOPmin}$ and $\CUSTOMerrof{\CUSTOMbetamin}$ as only one final value is calculated through \cref{eq:appendix:hist_rew_dlnm_minimization} from all simulated data.
In order to overcome this problem, we can use a resampling technique.
We have chosen the \CUSTOMdefhighlight{jackknife resampling} technique which is described, \CUSTOMeg{} in \autocite{shao1995} in full length.
We will only sketch the main steps applied in this work.
As our measurements were two-dimensional arrays consisting of time series $i = 1, \dots, \CUSTOMnmeas$ and disorder realizations $c = 1, \dots, \CUSTOMncfg$, we apply the resampling in both directions separately and combine the estimates at the end.
For each jackknife resampling step $\CUSTOMjki$ in the time series direction we leave out a block $\CUSTOMJK^{\CUSTOMjki} \subset \CUSTOMset{1, \dots, \CUSTOMnmeas}$ of measurements for each disorder realization $c$ so that the thermal average defined through \cref{eq:appendix:thermal_average_of_obs_O} becomes
\begin{align}
	(\CUSTOMO^c)^{\CUSTOMjki} = \CUSTOMoneover{\CUSTOMnmeas - \lvert\CUSTOMJK^{\CUSTOMjki}\rvert} \sum_{\substack{i=1 \\ i \notin \CUSTOMJK^{\CUSTOMjki}}}^{\CUSTOMnmeas} \CUSTOMO_i^c \;,
	\label{eq:appendix:thermal_average_jk}
\end{align}
where $\lvert\CUSTOMJK^{\CUSTOMjki}\rvert$ is the number of left-out samples.
Analogously, for each resampling step $\CUSTOMjkc$ in the disorder direction we leave out a block $\CUSTOMJK^{\CUSTOMjkc} \subset \CUSTOMset{1, \dots, \CUSTOMncfg}$ of disorder realizations so that the disorder average defined through \cref{eq:appendix:disorder_average_of_obs_O} becomes
\begin{align}
	(\CUSTOMO)^{\CUSTOMjkc} = \CUSTOMoneover{\CUSTOMncfg - \lvert\CUSTOMJK^{\CUSTOMjkc}\rvert} \sum_{\substack{c=1 \\ c \notin \CUSTOMJK^{\CUSTOMjkc}}}^{\CUSTOMncfg} \CUSTOMO^c \;.
	\label{eq:appendix:disorder_average_jk}
\end{align}
where $\lvert\CUSTOMJK^{\CUSTOMjkc}\rvert$ is the number of left-out realizations.

Starting from the modified thermal averages $(\CUSTOMO^c)^{\CUSTOMjki}$ and disorder averages $(\CUSTOMO)^{\CUSTOMjkc}$ respectively, all steps in the following analysis remain the same.
Let $\CUSTOMOA$ be a final estimate coming from a certain analysis, \CUSTOMeg{} minimum search as in \cref{eq:appendix:hist_rew_dlnm_minimization}.
By repeating the analysis for $\CUSTOMNjki$ different jackknife blocks in the time direction and $\CUSTOMNjkc$ blocks in the disorder direction we are given two arrays of estimates $(\CUSTOMOA)^{\CUSTOMjki}$ and $(\CUSTOMOA)^{\CUSTOMjkc}$ respectively.
We calculate two jackknife means $\CUSTOMmean{\CUSTOMOA}^{\CUSTOMjki}$ and $\CUSTOMmean{\CUSTOMOA}^{\CUSTOMjkc}$
\begin{align}
	\CUSTOMmean{\CUSTOMOA}^{\CUSTOMjka} & = \CUSTOMoneover{\CUSTOMNjka} \sum_{\CUSTOMjka=1}^{\CUSTOMNjka} (\CUSTOMOA)^{\CUSTOMjka} \CUSTOMwith \CUSTOMjka = \CUSTOMjki, \CUSTOMjkc  \;,
	\label{eq:appendix:jackknife_mean}
\end{align}
and two corresponding jackknife errors $\CUSTOMerrof{\CUSTOMOA}^{\CUSTOMjki}$ and $\CUSTOMerrof{\CUSTOMOA}^{\CUSTOMjkc}$
\begin{align}
	\CUSTOMerrof{\CUSTOMOA}^{\CUSTOMjka} & = \frac{\CUSTOMNjka - 1}{\CUSTOMNjka} \sum_{\CUSTOMjka=1}^{\CUSTOMNjka} \CUSTOMrbrl{(\CUSTOMOA)^{\CUSTOMjka} - \CUSTOMmean{\CUSTOMOA}^{\CUSTOMjka} }^2   \CUSTOMwith \CUSTOMjka = \CUSTOMjki, \CUSTOMjkc \;.
	\label{eq:appendix:jackknife_error}
\end{align}
As the last step we combine the two means and errors in a standard (uncorrelated) manner
\begin{align}
	\CUSTOMmean{\CUSTOMOA}  & = \CUSTOMhalf \CUSTOMrbrl{\CUSTOMmean{\CUSTOMOA}^{\CUSTOMjki} + \CUSTOMmean{\CUSTOMOA}^{\CUSTOMjkc}} \;                \\
	\CUSTOMerrof{\CUSTOMOA} & = \sqrt{\CUSTOMrbrl{\CUSTOMerrof{\CUSTOMOA}^{\CUSTOMjki}}^2 + \CUSTOMrbrl{\CUSTOMerrof{\CUSTOMOA}^{\CUSTOMjkc}}^2} \;.
	\label{eq:nu_exponent:final_hist_rew_minimal_mean_and_err}
\end{align}
The mean $\CUSTOMmean{\CUSTOMOA}$ and the corresponding error $\CUSTOMerrof{\CUSTOMOA}$ are the final results for a given analysis after applying jackknife resampling.
 
\bibliography{literature_bibtex} 

\end{document}